\documentclass[11pt]{article}
\usepackage{ifpdf}
\ifpdf
  \newif\ifhyperref \hyperreftrue
  \newif\ifadk \adkfalse
  \ifhyperref
  \usepackage[backref, citecolor=blue,%
    pdftitle={An Investigation of New Methods for Numerical Stochastic
    Perturbation Theory in $\varphi^4$ Theory},%
    pdfauthor={Mattia Dalla Brida, Marco Garofalo, A.~D.~Kennedy},%
    pdfkeywords={NSPT, ISPT, HSPT, KSPT, Stochastic Perturbation
      Theory, Quantum Field Theory},%
    \ifadk
    pdfstartview={FitH},%
    pdfpagescrop={60 50 550 730},
    pdfpagemode={FullScreen}%
    \fi
    ]{hyperref}
  \fi
\fi
\usepackage[english]{babel}
\usepackage{simplewick}
\usepackage{amsfonts}
\usepackage{amsmath}
\usepackage{amssymb}
\usepackage{graphicx}
\usepackage{booktabs}
\usepackage{sint}
\usepackage{xcolor}
\usepackage{pifont}
\graphicspath{{figures/}}
\def\rmd{{\rm d}}

\def\proof{\noindent{\sl Proof:}\kern0.6em}

\def\dual{\mathstrut^*\kern-0.1em}

\def\lvec#1{\setbox0=\hbox{$#1$}
    \setbox1=\hbox{$\scriptstyle\leftarrow$}
    #1\kern-\wd0\smash{
    \raise\ht0\hbox{$\raise1pt\hbox{$\scriptstyle\leftarrow$}$}}
    \kern-\wd1\kern\wd0}
\def\rvec#1{\setbox0=\hbox{$#1$}
    \setbox1=\hbox{$\scriptstyle\rightarrow$}
    #1\kern-\wd0\smash{
    \raise\ht0\hbox{$\raise1pt\hbox{$\scriptstyle\rightarrow$}$}}
    \kern-\wd1\kern\wd0}
\def\cvec#1{\kern-0.5pt\vec{\kern0.5pt #1}}

\def\slash#1{\setbox2=\hbox{$\displaystyle#1$}%
             \setbox3=\hbox{$\displaystyle/$}%
             #1\kern-0.8\wd2/\kern-1.0\wd3\kern0.8\wd2\kern0.5pt}

\def\wick#1{\setbox2=\hbox{$\displaystyle#1$}
    \setbox3=\null\ht3=3.0pt\dp3=0.0pt\wd3=20.0pt
    #1\kern-\wd2\kern3.0pt\raise11.0pt\vbox{\hrule height0.3pt
    \hbox{\vrule width0.3pt\box3\vrule width0.3pt}}\kern-24.0pt\kern\wd2}

\def\longwick#1{\setbox2=\hbox{$\displaystyle#1$}
    \setbox3=\null\ht3=3.0pt\dp3=0.0pt\wd3=27.0pt
    #1\kern-\wd2\kern3.0pt\raise11.0pt\vbox{\hrule height0.3pt
    \hbox{\vrule width0.3pt\box3\vrule width0.3pt}}\kern-31.0pt\kern\wd2}

\def\verylongwick#1{\setbox2=\hbox{$\displaystyle#1$}
    \setbox3=\null\ht3=3.0pt\dp3=0.0pt\wd3=43.0pt
    #1\kern-\wd2\kern3.0pt\raise11.0pt\vbox{\hrule height0.3pt
    \hbox{\vrule width0.3pt\box3\vrule width0.3pt}}\kern-47.0pt\kern\wd2}

\def\nabstar#1{{\nabla\kern0.5pt\smash{\raise 4.5pt\hbox{$\ast$}}
               \kern-5.5pt_{#1}}}
\def\drv#1{{\partial_{#1}}}
\def\drvstar#1{{\partial\kern0.5pt\smash{\raise 4.5pt\hbox{$\ast$}}
               \kern-6.0pt_{#1}}}
\def\sdrvstar#1{{\partial\kern0.4pt\smash{\raise 3.6pt\hbox{$\ast$}}
                \kern-4.8pt_{#1}}}

\def\ldrvstar#1{{\lvec{\,\partial}\kern-0.5pt\smash{\raise 4.5pt\hbox{$\ast$}}
               \kern-5.0pt_{#1}}}

\def\MSbar{\overline{\rm MS\kern-0.5pt}\kern0.5pt}

\def\diracstar#1#2{
    \setbox0=\hbox{$\gamma$}\setbox1=\hbox{$\gamma_{#1}$}
    \gamma_{#1}\kern-\wd1\kern\wd0
    \smash{\raise4.5pt\hbox{$\scriptstyle#2$}}}

\def\Dw{D_{\rm w}}
\def\Dwdag{{\Dw}\kern-4pt^{\dagger}\kern1pt}
\def\Dm{D}
\def\Dmdag{\Dm^{\dagger}\kern-1pt}

\def\Dmsdag{D_s\kern-2pt\vphantom{D}^{\dagger}\kern-1pt}

\def\sump#1#2{\sum_{#1}^{#2}\kern-2pt{\vphantom{\sum}}'}
\def\Ebar{\kern1.5pt\overline{\kern-1.5ptE\kern-0.5pt}\kern0.5pt}
\def\obs{{\cal O}}
\newcommand{\order}{O}	
\newcommand{\Z}{\mathbb{Z}}     

\newif\ifcomment
 \commenttrue

\begin{document}
%

\begin{titlepage}
  \hfill CERN-TH-2017-059  
  
  \null\vspace*{15mm}
  \begin{center}
    {\Large\bf Investigation of New Methods for Numerical Stochastic
      Perturbation Theory in \(\varphi^4\) Theory}
  \end{center}
  \vspace{0.75cm}
  \begin{center}
    {\large Mattia Dalla Brida\(^{\scriptscriptstyle a}\),
      Marco Garofalo\(^{\scriptscriptstyle b}\), and
      A.~D.~Kennedy\(^{\scriptscriptstyle b}\) \\}
    \vspace{0.75cm} \(^{\scriptstyle a}\) Dipartimento di Fisica, Universit\`a
    di Milano-Bicocca and \\ INFN, Sezione di Milano-Bicocca, Piazza della
    Scienza 3, I-20126 Milano, Italy \\
    \vspace{1.5ex} \(^{\scriptstyle b}\) Higgs Centre for Theoretical Physics,
    School of Physics and Astronomy, \\ The University of Edinburgh, Edinburgh
    EH9 3FD, Scotland, United Kingdom \\

    \vspace{2.0cm} {\bf Abstract}
    \vspace{0.35ex}
  \end{center}
  {\noindent Numerical stochastic perturbation theory is a powerful tool for
    estimating high-order perturbative expansions in lattice field theory. The
    standard algorithms based on the Langevin equation, however, suffer from
    several limitations which in practice restrict the potential of this
    technique.  In this work we investigate some alternative methods which
    could in principle improve on the standard approach. In particular, we
    present a study of the recently proposed Instantaneous Stochastic
    Perturbation Theory, as well as a formulation of numerical stochastic
    perturbation theory based on Generalized Hybrid Molecular Dynamics
    algorithms.  The viability of these methods is investigated in
    \(\varphi^4\) theory.\parfillskip=0pt\par} \vfill\eject

\end{titlepage}



\section{Introduction}

Lattice perturbation theory (LPT) is an important tool in lattice field theory,
and in particular in related renormalization problems (see,
e.g.,~\cite{Rothe:1992nt,Montvay:1994cy,Capitani:2002mp} for an introduction).
LPT may be used to compute the matching of physical renormalization schemes
employed on the lattice and schemes commonly used in continuum perturbative
calculations, such as the \(\overline{\rm MS}\)-scheme of dimensional
regularization.\footnote{Physical renormalization schemes are those that do not
  explicitly depend on the regulator.} In addition LPT gives insight into
lattice artefacts of the theory, allowing for both the perturbative
determination of Symanzik improvement coefficients and, more generally, of the
lattice artefacts in observables of interest.

LPT is technically much more involved than its continuum counterpart because of
the complicated form of its vertices and propagators, and usually requires
numerical evaluation for even simple diagrams. This is especially true when
sophisticated lattice discretizations are considered. Additionally, in the case
of gauge theories, the appearance of new vertices at every order of
perturbation theory makes the number of diagrams grow very rapidly with the
perturbative order, leaving only low-order results accessible to standard
techniques.

Numerical stochastic perturbation theory (NSPT) was proposed long
ago~\cite{DiRenzo:1994av,DiRenzo:1994sy} (see~\cite{DiRenzo:2004ge} for a
detailed review, and~\cite{Hesse:2013lat,Brida:2013mva,Bali:2013pla} for recent
developments) in order to circumvent these general difficulties, and thus
enable high-order perturbative computations in LPT. The basic idea of NSPT is
the numerical integration of a discrete version of the equations of stochastic
perturbation theory~\cite{Parisi:1980ys} (see~\cite{Damgaard:1987rr} for a
review). More precisely, starting from the Langevin equation the stochastic
field is expanded as a power series in the couplings of the theory and the
resulting equations are solved order by order in these couplings. No Feynman
diagrams need to be identified or computed, but rather a system of stochastic
differential equations is integrated numerically using Monte Carlo techniques. 
In this framework perturbative calculations may be highly automated. Complicated
observables can be considered with no additional difficulty, and the cost of
these methods scales mildly with the perturbative order. In principle, NSPT
allows high-order perturbative determinations even in cases where the
corresponding continuum calculations are not feasible.

Of course this requires that the continuum limit can be evaluated reliably.
This is a limitation that may restrict the applicability of NSPT. Firstly, the
results at finite lattice resolution unavoidably come with statistical
uncertainties due to their Monte Carlo estimation. In particular, the numerical
simulations suffer from critical slowing down as the continuum limit of the
theory is approached; this significantly increases the computational effort
necessary to extract continuum results from NSPT. Secondly, this class of
algorithms is not exact: therefore a sequence of simulations with finer and
finer discretization of the relevant equations must be performed in order to
extrapolate away systematic errors in the results. It is thus difficult to
obtain precise results close to the continuum limit for which both systematic
and statistical errors are under control. Without continuum extrapolation these
methods only provide lattice estimates for perturbative quantities, which in
practice may be of limited use.

Experience with conventional algorithms for non-perturbative lattice field
theory simulations suggests that a different choice of stochastic process might
significantly alleviate these limitations. In particular, the class of methods
known as Generalized Hybrid Molecular Dynamics (GHMD) algorithms have proven to
be superior to Langevin algorithms in this respect; in fact the latter are a
special case of the former.

From a different perspective Martin L\"uscher recently introduced a new form of
NSPT, namely Instantaneous Stochastic Perturbation Theory
(ISPT)~\cite{Luscher:2014mka}.  In this work, he discussed how the above
limitations can in principle be eliminated completely by formulating NSPT in
terms of a certain class of trivializing fields.  This method lies somewhere
between Langevin NSPT and more conventional diagrammatic perturbation theory.

The aim of this work is to compare the standard NSPT formulation, ISPT, and
NSPT based upon GHMD algorithms. Specifically, we will focus on two GHMD
algorithms, namely the Hybrid Molecular Dynamics (HMD) algorithm and Kramers
algorithm.

The structure of the paper is as follows. In \S\ref{sec:definitions} we give
some general definition including the lattice action and observables used in
this study. In \S\ref{sec:ISPT} we review ISPT, paying attention to its
numerical implementation. \S\ref{sec:LSPT} is dedicated to a review of the
standard NSPT approach based on the Langevin equation (LSPT). In
\S\ref{sec:GHMD} we introduce NSPT based on the HMD algorithm (HSPT) and
Kramers algorithm (KSPT). Finally, in \S\ref{sec:numres} we present results of
the numerical investigation of the different methods, followed by our
conclusions.  Preliminary results of our study appeared
in~\cite{Brida:2015vfv}.



\section{Definitions} \label{sec:definitions}

\subsection{Lattice theory}

We consider the simple \(\varphi^4\) theory, with \(\varphi\) a single
component real field, defined on a four-dimensional Euclidean lattice of extent
\(L\) in all directions.  The theory is specified by the lattice action,
\begin{equation}
  \label{eq:S}
  S(\varphi)=
  a^4\sum_{x\in\Omega}\,\left(\frac12 \drv\mu\varphi(x) \drv\mu\varphi(x)
  + \frac12 m^2_0\varphi(x)^2 + {\frac{g_0}{4!}}\varphi(x)^4\right),
\end{equation}
where \(\varphi\) is the \emph{bare} field, \(\partial_\mu \varphi(x) =
\bigl(\varphi(x+a\hat\mu)-\varphi(x)\bigr)/a\) is the usual forward lattice
derivative with \(\hat\mu\) being a unit vector in the direction
\(\mu=0,\ldots,3\), and \(a\) is the lattice spacing.  The sum in (\ref{eq:S})
runs over the set \(\Omega\) of all lattice points \(x=(x_0,x_1,x_2,x_3)\) with
\(x_i/a\in \Z_{L/a}\), while the field \(\varphi\) satisfies the periodicity
conditions \(\varphi(x+\hat{\mu}L)= \varphi(x)\), \(\forall \mu\). The
parameters \(m_0\) and \(g_0\) are the \emph{bare} mass and coupling
constant; they are related to the renormalized quantities \(m\) and \(g\)~by
\begin{align}
  \label{eq:dm2}
  m^2 &= m_0^2 - \delta m^2=m^2_0 - \sum_{k=1}^\infty m^2_k\,g_0^k, \\
  \label{eq:dg}
  g &= g_0 - \delta g = g_0 + \sum_{k=2}^\infty c_k\,g_0^{k},
\end{align}
where the coefficients \(m^2_k\) and \(c_k\) of the mass and coupling
counterterms \(\delta m^2\) and \(\delta g\) are determined order by order in
the coupling from the renormalization conditions; these are discussed below.

Given these definitions the expectation value of a generic observable
\(\obs(\varphi)\) of the field is defined as usual through the Euclidean
functional integral
\begin{equation}
 \label{eq:ExpectationValue}
 \langle\obs\rangle
 = \frac1{\mathcal{Z}}\int D\varphi\,e^{-S(\varphi)}\obs(\varphi), \quad
 D\varphi\equiv\prod_{x\in\Omega}\rmd\varphi(x),
\end{equation}
where the constant \(\mathcal{Z}\) is fixed by the condition \(\langle1\rangle
= 1\).  Of interest for the following discussion is the bare two-point
function,
\begin{equation}
  \chi_2(p) = a^4\sum_{x\in\Omega} e^{-ipx}
    \bigl\langle\varphi(x)\varphi(0)\bigr\rangle,
\end{equation}
where \(p=(p_0,p_1,p_2,p_3)\) with \(p_i=2\pi n_i/L\) and \(n_i\in\Z_{L/a}\)
are the allowed momenta in a periodic box; the set of such momenta will be
denoted in the following by~\({\tilde\Omega}\). In particular, we will consider
\begin{equation}
 \chi_2\equiv\chi_2(0)\quad{\rm and}\quad
 \chi_2^*\equiv\chi_2(p_*),
\end{equation}
where \(p_*\) is the minimal non-zero momentum given by \(p_{*} =
(2\pi/L,0,0,0)\).\footnote{In general we shall consider lattice units where
  \(a=1\) from now on. Nevertheless, the lattice spacing may be included in
  some formulas for clarity.}

\subsection{Renormalization conditions and observables}

In order to study the continuum limit of the theory some renormalization
conditions must be chosen to define the renormalized parameters and fields; we
use the finite size renormalization scheme described in~\cite{Weisz:2010xx}.
For simplicity we study the symmetric phase of the theory, although the methods
we shall present can be adapted to the spontaneously broken phase too.

Our definition of a renormalized mass \(m\) is obtained from
\begin{equation}
 \label{eq:MassDefinition}
 \frac{\chi_2}{\chi^*_2} = 1+\frac{{\hat p_*}^2}{m^2},
\end{equation}
where \({\hat p}^2=\sum_\mu {\hat p}^2_\mu\), with \(\hat{p}_\mu =
2\sin(p_\mu/2)\) being the usual lattice momenta. The finite size continuum
limit may then be defined by keeping the combination
\begin{equation}
  z=mL
\end{equation}
fixed.

More precisely, for a given choice of \(z\) the continuum limit is approached
by taking the lattice size \(L = L/a\to\infty\) while tuning the lattice mass
\(m = am\to0\), such that \(z\) has the desired value. The possible values of
\(z\) thus identify a \emph{family} of renormalization schemes.

The wavefunction renormalization \(Z = Z(g_0,L/a,am)\) which defines the
renormalized elementary field \(\varphi_R(x)=Z^{-1/2} \varphi(x)\) is fixed~by
\begin{equation}\label{eq:Z_def}
  Z^{-1} = \frac{\chi_2^{*\,-1} - \chi_2^{-1}}{{\hat p_*}^2}
  \quad\Longrightarrow\quad Z = m^2\chi_2.
\end{equation}
Given these definitions, we introduce the renormalized coupling
\begin{equation}
  \label{eq:gR}
  g = -\frac{\chi_4}{\chi_2^2}\,m^4,
\end{equation}
where \(\chi_4\) is the bare connected four-point function at zero external
momenta,
\begin{equation}
 \label{eq:chi4}
 \chi_4 = \sum_{x,y,z\in\Omega}
   \bigl\langle\varphi(x)\varphi(y)\varphi(z)\varphi(0)\bigr\rangle
   - 3L^4 \chi_2^2.
\end{equation}
The above renormalization conditions are a natural
extension of textbook renormalization conditions for \(\varphi^4\) theory in a
finite lattice volume. What is relevant for the present study is the fact that
the coupling (\ref{eq:gR}) is known to two-loop order in lattice perturbation
theory~\cite{Weisz:2010xx}: this provides us with a non-trivial result to
compare with. On the other hand, a precise determination of (\ref{eq:chi4})
using the Monte Carlo methods presented in the next sections is difficult on
large lattices (required to be close to the continuum limit) due to the
stochastic subtraction of the disconnected contribution.

In order to obtain precise and simple quantities with well-defined continuum
limits we consider observables defined through the gradient flow
(see~\cite{Luscher:2010iy,Luscher:2011bx} for an introduction). In the case of
the \(\varphi^4\) theory the gradient flow equations take the simple
form~\cite{Monahan:2015lha,Monahan:2015fjf}
\begin{equation}
 \label{eq:GF}
 \partial_t{\tilde\varphi}(t,x)
 = \partial^2\tilde{\varphi}(t,x)\quad\mbox{with}\quad
   \tilde{\varphi}(0,x)=\varphi_R(x),
\end{equation}
where \(t\geq0\) is the flow time and
\(\partial^2=\sum_\mu\drvstar\mu\drv\mu\), with \(\drvstar\mu\varphi(x) =
\varphi(x)-\varphi(x-{\hat\mu})\), is the usual lattice Laplacian. In
particular, products of fields at positive flow time are automatically
renormalized if the parameters of the theory are renormalized.  The
dimensionless quantity
\begin{equation}
  \label{eq:Eoft}
  {\cal E}(t) = t^2 \langle E(t,x)\rangle \quad\mbox{with}\quad
  E(t,x) = \tilde\varphi(t,x)^4,
\end{equation}
for example, is finite without any additional renormalization, provided that
the physical flow time \(t\) is held fixed as the continuum limit of the theory
is approached.  Hence, we define the finite size continuum limit of flow 
quantities like (\ref{eq:Eoft}) by holding the ratio~\cite{Fodor:2012td}
\begin{equation}
  c=\sqrt{8t}/L
\end{equation}
fixed.  The continuum limit is thus taken by increasing the lattice size \(L =
L/a\) and the flow time in lattice units \(t= t/a^2\) such that \(c\) is
fixed to some chosen value; different values of \(c\) define different
renormalization schemes.



\section{An implementation of ISPT in \(\varphi^4\) theory}
\label{sec:ISPT}

The first new technique we present is ISPT. Here we limit ourselves to
describing the essential features of this approach in order to emphasize the most
prominent differences with standard NSPT techniques. This short review will
also help introduce our notation and some concepts useful for later
discussions. We recommend the reader to the original
reference~\cite{Luscher:2014mka} where a detailed presentation is to be
found.\footnote{Additional useful material is provided by the author
  of~\cite{Luscher:2014mka} in the documentation for the publicly available
  package~\cite{LuscherWeb:2014}.}

\subsection{Definitions}

ISPT is based on the concept of trivializing maps. In the most general case
these transform a set of Gaussian-distributed random fields \(\eta_{i}(x)\),
for \(i=0,1,2,\ldots\), into a stochastic field \(\phi(x)\) such that
\begin{equation}
 \label{eq:Trivialization}
 \bigl\langle \phi(x_1)\cdots\phi(x_n)\bigr\rangle_\eta
 = \bigl\langle \varphi(x_1)\cdots\varphi(x_n)\bigr\rangle
\end{equation}
order by order in the couplings of the theory. Here the expectation value on
the right hand side is defined by~(\ref{eq:ExpectationValue}), whereas that on
the left hand side it is given in terms of averages over the Gaussian random fields:
\begin{equation}
 \bigl\langle\eta_i(x)\bigr\rangle_\eta = 0,\qquad 
 \bigl\langle\eta_i(x)\eta_j(y)\bigr\rangle_\eta = \delta_{ij}\delta_{xy}.
\end{equation}
In perturbation theory the stochastic field \(\phi\) can be represented as a
power series in the couplings of the theory. In particular, in the regularized
theory we can consider an expansion in terms of the bare coupling \(g_0\),
\begin{equation}
 \label{eq:PowerSeriesField}
 \phi(x) =\sum_{k=0}^{N} \phi_{k}(x)g_0^k + \order{(g_0^{N+1})}.
\end{equation}
If this is given the corresponding expansion in terms of a renormalized
coupling is easily obtained using relation (\ref{eq:dg}) (q.v.,
Appendix~\ref{subsec:CouplingRenormalization}). On the other hand, the
determination of the coefficients \(\phi_k\) in terms of the renormalized mass,
instead of the bare mass, requires explicit computation of the mass counterterm
contributions. For the numerical implementation of the method it is thus
convenient to store the field as a two-dimensional array \(\phi_{k,\ell}\) with
the indices corresponding to the powers of \(g_0\) and~\(\delta m^2\):
\begin{equation}
 \label{eq:DoublePowerSeriesField}
 \phi(x) = \sum_{k,\ell=0}^{N} \phi_{k,\ell}(x)g_0^k(\delta m^2)^\ell
   + \order{(g_0^{N+1})}.
\end{equation}
Once the expansion (\ref{eq:dm2}) is known it is trivial to pass from the
representation (\ref{eq:DoublePowerSeriesField})
to~(\ref{eq:PowerSeriesField}).  Using the representation
(\ref{eq:DoublePowerSeriesField}) the expansion (\ref{eq:dm2}) can be
determined and thus the results obtained in terms of the renormalized mass.
This is discussed in detail in Appendix~\ref{subsec:MassRenormalization}; we
recommend that the reader consults this appendix only after reading the remainder
of this section in which all the relevant definitions are introduced.

We find at the lowest-order in the coupling
\begin{equation}
  \label{eq:LowestOrderField}
  \phi_{0,0}(x) = \sum_{y\in\Omega} H(x,y)\eta_0(y),
\end{equation}
where \(H\) is the Green function for the operator \(\sqrt{-\partial^2+m^2}\),
\begin{equation}
  H(x,y) = \frac{1}{L^4}\sum_{p\in{\tilde \Omega}}e^{ip(x-y)}\sqrt{\tilde G(p)},
  \qquad\mbox{where}\qquad {\tilde G}(p)=\frac{1}{{\hat p}^2+m^2}.
\end{equation}
It is easy to show that this field satisfies~(\ref{eq:Trivialization}) at
lowest order in the coupling.

\begin{figure}[hpbt]
 \centering
 \includegraphics[scale=0.7]{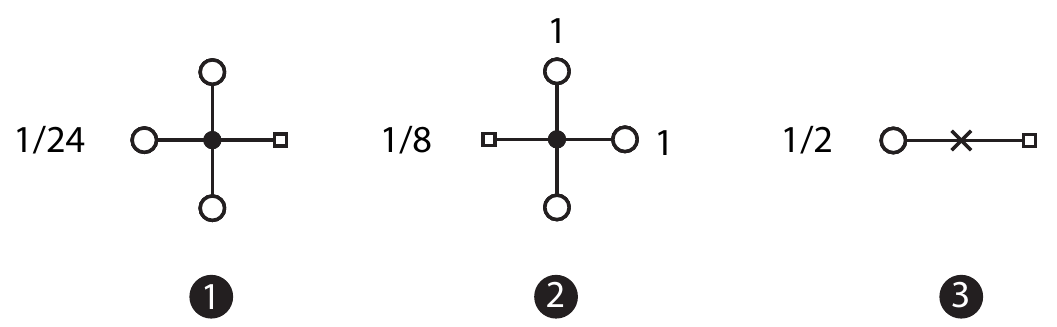}
 \caption{Rooted tree-diagrams contributing at \(\order(g_0)\); note that
   \(\delta m^2 = \order(g_0)\).}
 \label{fig:RootedTreesScalar1}
 \includegraphics[scale=0.7]{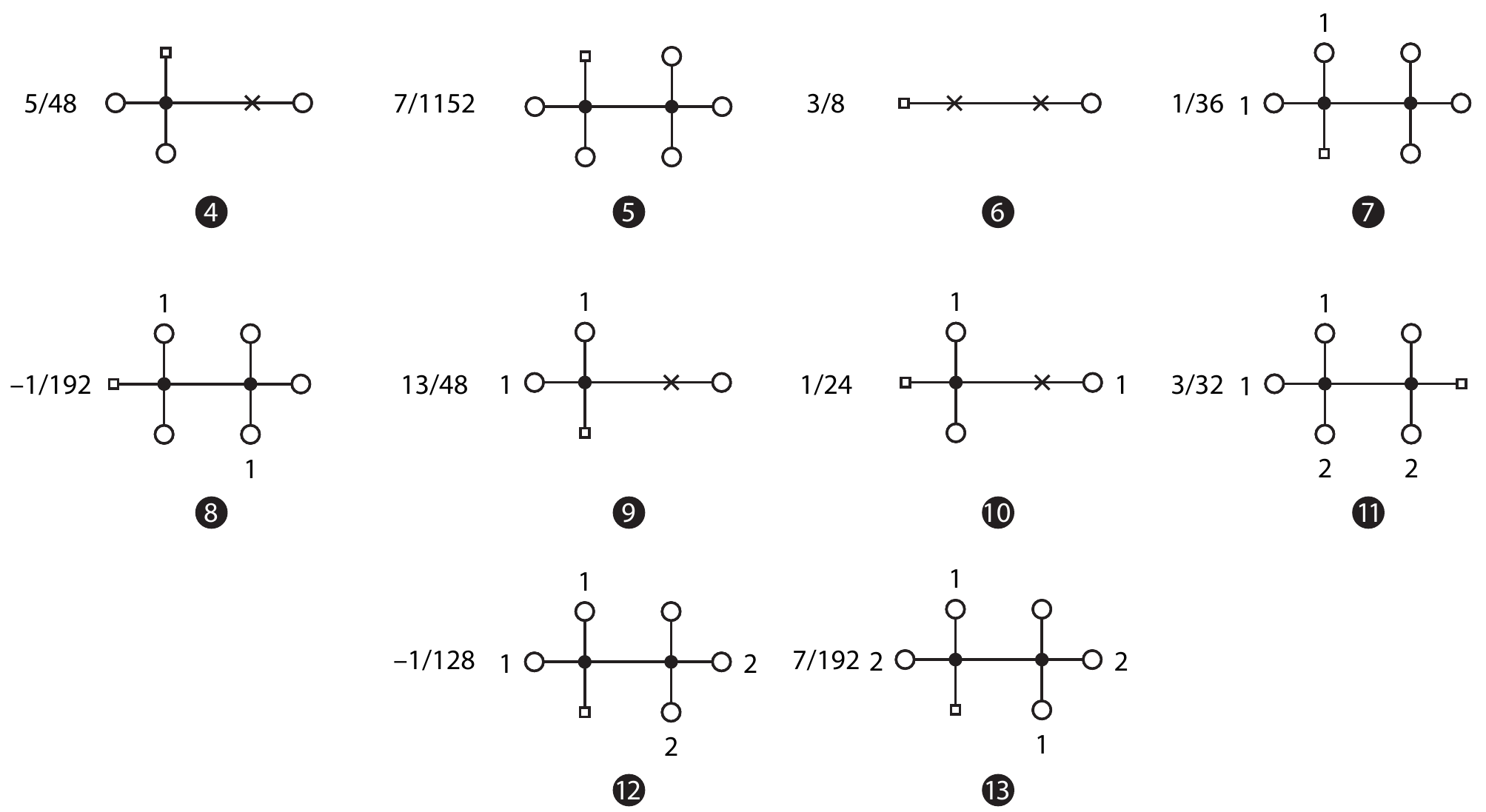}
 \caption{Rooted tree-diagrams contributing at \(\order(g_0^2)\); note that
   \(\delta m^2 = \order(g_0)\).}
 \label{fig:RootedTreesScalar2}
\end{figure}

Beyond the leading order there is more freedom to define the trivializing
field. Following~\cite{Luscher:2014mka} we write this as a linear combination
of the values \(v(x,\mathcal{R}_i)\) of the rooted tree diagrams
\(\mathcal{R}_i\) with coefficients~\(c(\mathcal{R}_i)\),
\begin{equation}
 \phi_{k,l}(x)=\sum_{i\in \mathcal{S}_{k,l}} c(\mathcal{R}_i) v(x,\mathcal{R}_i),
\end{equation}
where \(\mathcal{S}_{k,l}\) is the set of all diagrams of order \(g_0^k\) and
\((\delta m^2)^l\).  Graphical representations of the rooted tree-diagrams
contributing to \(\order(g_0)\) (\(k+\ell = 1\)) and \(\order(g_0^2)\)
(\(k+\ell= 2\)) are given in Figures~\ref{fig:RootedTreesScalar1}
and~\ref{fig:RootedTreesScalar2} respectively; the corresponding coefficients
\(c(\mathcal{R}_i)\) are also shown. In this representation the leaves of the
trees are given by
\begin{equation}
  \includegraphics[scale=0.15, trim=0 27 0 0]{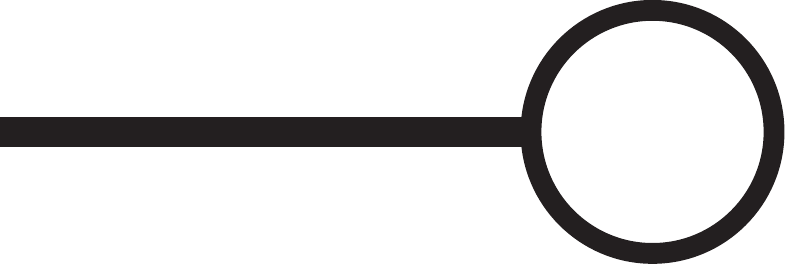} = 
  \chi_i(x) = \sum_{y\in\Omega} H(x,y) \eta_i(y),
\end{equation}
where the index \(i\) is the number adjacent to the open circle in the graph;
if no such number is displayed it is implicit that~\(i=0\). The leaves are thus
given by the lowest order solution~(\ref{eq:LowestOrderField}) with the
appropriate choice of random field~\(\eta_i\).

Black circles and crosses represent the vertices of the theory: they are the
usual \(\varphi^4\) vertex and mass counterterm insertions,
\begin{equation}
  \includegraphics[scale=0.3, trim=0 35 0 0]{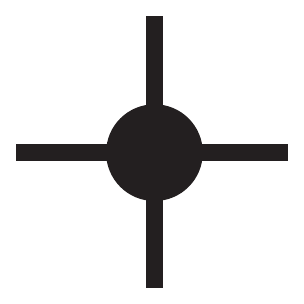} = -1,\qquad
  \includegraphics[scale=0.2, trim=0 20 0 0]{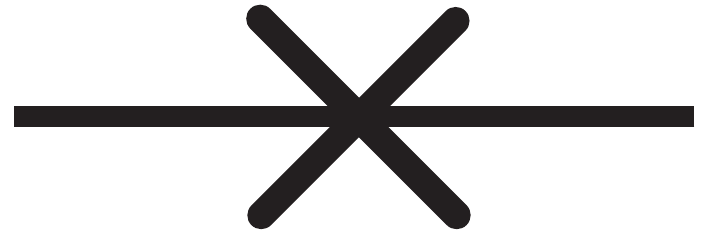} = 1.
\end{equation}
These are associated with implicit factors of \(-g\) and \(\delta m^2\)
respectively. Black lines connecting two vertices correspond to the scalar
propagator,
\begin{equation}
  \includegraphics[scale=0.2, trim=0 -3 0 0]{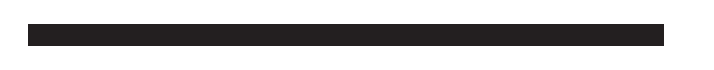} = G(x,y)
  =\frac1{L^4}\sum_{p\in{\tilde \Omega}}e^{ip(x-y)}{\tilde G}(p),
 \label{eq:ScalarPropagator}
\end{equation}
where \(x\) and \(y\) are the positions of the two vertices connected by the
given propagator. In particular, at each vertex the fields attached are
multiplied together and the propagator is applied to the resulting product of
fields.

The root of the diagram is given by
\begin{equation}
  \includegraphics[scale=0.25, angle=180, trim=0 38 0 0]{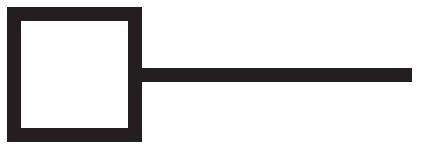} = G(x,y),
\end{equation}
where \(x\) is the space-time index of the corresponding rooted
tree~\(\mathcal{R}_i\).

To give some examples, given some \(\eta_0(x)\) and \(\eta_1(x)\) fields, the
diagram labeled~\ding{183} in Figure~\ref{fig:RootedTreesScalar1} evaluates~to
\begin{equation}
  v(x,\mathcal{R}_2)=(-1)\,\sum_{y\in\Omega} G(x,y)\chi_0(y)\chi_1(y)^2.
\end{equation}
This contributes to \(\phi_{1,0}(x)\) with a
coefficient~\(c(\mathcal{R}_2)=1/8\).  Diagram \ding{185} in
Figure~\ref{fig:RootedTreesScalar2}, stands for
\begin{equation}
  v(x,\mathcal{R}_4)=(-1)\,\sum_{y\in\Omega} G(x,y)\chi_0(y)^2
  \sum_{z\in\Omega} G(y,z)\chi_0(z),
\end{equation}
and contributes to \(\phi_{1,1}(x)\) with \(c(\mathcal{R}_4)=5/48\).

\begin{table}[hbpt]
 \centering
 \begin{tabular}{rrr}
  \toprule
  \(g_0^k\)      &  \(n\)  & \(c(\mathcal{R})\neq0\) \\
  \midrule
   1 &      3 &      3 \\
   2 &     10 &     10 \\
   3 &     44 &     43 \\
   4 &    241 &    231 \\
   5 &  1,506 &  1,420 \\
   6 & 10,778 & 10,015 \\
   \midrule
   Total & 12,582 & 11,722 \\
   \bottomrule       
 \end{tabular}
 \caption{Number of rooted-tree diagrams appearing at a given order in the
   coupling \(g_0\). The column labeled by \(c(\mathcal{R})\neq0\) gives the
   number of such diagrams whose coefficient \(c(\mathcal{R}_i)\) is
   non-vanishing.}
 \label{tab:NumberDiagrams}
\end{table}

Given these examples it is clear that the evaluation of the trivializing map
for a given set of random fields \(\eta_i\) is in principle straightforward.
Beyond the lowest perturbative orders though the number of diagrams (as well
as their complexity) increases rapidly as indicated in Table~\ref{tab:NumberDiagrams},
so the computation must be automated.

For this work we wrote a program that evaluates the trivializing field
\(\phi(x)\) up to an arbitrary order \(N\) in the couplings for a given set
of \(\eta_i\) fields. For the structure of the relevant diagrams and the
determination of their coefficients we used the software package provided by
Martin L\"uscher~\cite{LuscherWeb:2014}. The diagrams are given as {\tt
  C~structs} of abstract elements, so our program visits each vertex in a
diagram using depth-first recursion starting from the root, and evaluates the
corresponding numerical expressions.  The diagrams are collected according to
their order in the couplings and the \(\phi_{k,l}(x)\) fields are thus
constructed. This allows the series (\ref{eq:DoublePowerSeriesField}) to be
obtained for some set of \(\eta_i\) fields. Once this is done, correlation
functions of the trivializing field can be expanded order by order in the
couplings and evaluated stochastically by averaging over different samples of
the Gaussian random fields~\(\eta_i\). In particular, the perturbative
expansion of generic observables of the trivializing field \(\obs(\phi)\) can
be computed by iterating order by order convolution operations of the
form,
\begin{equation}
  \label{eq:OrderByOrderProduct}
  (\phi\cdot\phi)(x,y) = \phi(x)\phi(y)\quad\Longrightarrow\quad
  (\phi\cdot\phi)_{k,\ell}(x,y) = \sum_{0\leq i\leq k}\sum_{0\leq j\leq\ell}
    \phi_{k-i,\ell-j}(x)\phi_{i,j}(y),
\end{equation}
and similarly for other elementary operations. In this way one obtains the
generic stochastic perturbative field,
\begin{equation}
 \label{eq:StochasticObservable}
 \obs(\phi) = \sum_{k,\ell=0}^N \obs_{k,\ell}(\phi_{0,0},\ldots,\phi_{k,\ell})
   \,g_0^k(\delta m^2)^\ell + \order(g_0^{N+1}),
\end{equation}
from which the perturbative expansion of the expectation value of the field
\(\obs(\varphi)\) in \(\varphi^4\)~theory,
\begin{equation}
  \label{eq:PerturbativeVEV}
  \langle\obs\rangle =\sum_{k,\ell=0}^N a_{k,\ell}\, g_0^k(\delta m^2)^\ell
    + \order(g_0^{N+1}),
\end{equation}
is obtained up to \(\order(g_0^{N+1})\) corrections as
\begin{equation}
  \label{eq:PowerSeriesGreenFunction}
  \langle\obs\rangle_\eta = \langle\obs\rangle
  \quad\iff\quad\langle\obs_{k,\ell}\rangle_\eta = a_{k,\ell}.
\end{equation}
Once the expansion (\ref{eq:StochasticObservable}) is known the corresponding
expansion in terms of a given renormalized mass and coupling (as well as any
renormalization of the field \(\obs\)) is easily found (q.v.,
Appendix~\ref{sec:appA}).

We should mention some additional technical details.  First, in the
diagrammatic computation the scalar propagators are applied in momentum space,
while the products of fields at vertices are performed in position space.  This
is implemented using the efficient numerical evaluation of the discrete Fourier
transformation provided by the FFTW package~\cite{FFTW05}.  As a result the
cost of the computation of the diagrams scales proportionally to the system
size \(V=L^4\) up to logarithms.  Second, as already noted
in~\cite{Luscher:2014mka}, the computation of the rooted tree diagrams could be
organized in such a way that identical sub-trees in different graphs are
cached.  How to do this efficiently is a non-trivial issue even for
\(\varphi^4\) theory, and we did not investigate it further.  Moreover, whether
this is really worth investigating is not clear since, as we shall see below,
ISPT suffers from some severe limitations once high-order computations are
considered. Its utility might thus be limited to relatively low-order
computations where recomputation of subgraphs is not a significant issue.

The advantages of ISPT are that its results are exact up to statistical
uncertainties and that there are no autocorrelations as the coefficients
\(\phi_{k,\ell}\) are generated ``instantaneously'' from independent Gaussian
random fields~\(\eta_i\).

\subsection{A test of the method}

We tested our ISPT implementation by comparing some results with those obtained
using conventional perturbative lattice calculations (LPT).  We computed the
renormalized coupling (\ref{eq:gR}) and compared it with its two-loop
determination from~\cite{Weisz:2010xx}, which we evaluated for the parameters
of interest (see below).  We considered both the case where the perturbative
expansion is given in terms of the renormalized mass (\ref{eq:MassDefinition}),
and the case where it is given in terms of the bare mass \(m_0\).\footnote{In
  ISPT the latter is simply obtained by setting \(\delta m^2=0\) in the
  corresponding expansion~(\ref{eq:StochasticObservable}).}  The comparison was
done on a tiny lattice with \(L=4\), where high statistics could be gathered, 
and the value of the mass was chosen such that~\(z=4\). The results of the tests
are reported in Table~\ref{tab:AnalyticVsISPT}; for completeness we also give 
the results for \(\delta m^2\) in the~table.

\begin{table}[hbpt]
  \centering
  \begin{tabular}{cclllll}
  \toprule
  & Mass & \(c_2\times10^2\) & \(c_3\times10^3\) & \(m^2_1\times10^2\)
    & \(m^2_2\times10^4\)\\
  \midrule 
  LPT & \(m\) & \(-3.330\) & \(1.583\) & \(-6.4221\) & \(3.6702\) \\
  ISPT & \(m\) & \(-3.332(6)\) & \(1.582(4)\) & \(-6.4220(1)\)
    & \(3.6704(6)\) \\
  \midrule
  LPT & \(m_0\) & \(-3.33\) & \(2.965\) & & \\
  ISPT & \(m_0\) & \(-3.33(1)\)  & \(2.964(5)\) & & \\
  \bottomrule
  \end{tabular}
 \caption{Results for the series (\ref{eq:dg}) and (\ref{eq:dm2}) as obtained
   from ISPT and conventional LPT for \(L=4\) and \(z={\rm Mass}\times L\)
   using \(10^8\) field configurations. The perturbative expansion for the coupling
   (\ref{eq:dg}) is obtained both in terms of the renormalized mass \(m\)
   of~(\ref{eq:MassDefinition}) and the bare mass \(m_0\).}
 \label{tab:AnalyticVsISPT}
\end{table}

As can be seen from the table there is good agreement between the ISPT and the
LPT determinations, thus confirming the correctness of our implementation.  In
the case where the mass renormalization is considered one needs to take into
account the effect of statistical errors in the mass renormalization procedure
discussed in Appendix~\ref{subsec:MassRenormalization}: we did this using the
jackknife method.



\section{NSPT based on the Langevin equation}
\label{sec:LSPT}

Having introduced ISPT, in this and the following section we discuss the other
NSPT methods that we studied.  In these methods the stochastic field \(\phi\)
is generated through a Markov process based on some stochastic differential
equation expanded up to some fixed order in the couplings of the theory.  We
start from the standard NSPT based on the Langevin equation; for later convenience
we shall refer to this algorithm as LSPT.  This algorithm has a long history
and has been studied in great detail over the years: we thus limit ourselves 
to recalling the most relevant features for what follows, while referring the
reader to the literature for a more detailed account (see, e.g.,~\cite{DiRenzo:2004ge}
and references therein).

\subsection{Definition}
\label{subsec:LSPT}

The standard LSPT approach is based on stochastic
quantization~\cite{Parisi:1980ys,Floratos:1982xj,ZinnJustin:1986eq,
  ZinnJustin:1987ux,Damgaard:1987rr}, where the field representing the theory
is obtained as the solution of the Langevin equation,
\begin{equation}
  \label{eq:Langevin}
  \partial_{t_s} \phi(t_s,x) = -F\bigl(\phi(t_s,x)\bigr) +\eta(t_s,x),
\end{equation}
where \(F(\bigl(\phi(t_s,x)\bigr)\) denotes the functional derivative of the
action (\ref{eq:S}) evaluated on the field configuration \(\phi(t_s,x)\),
\begin{equation}
  \label{eq:Sder}
  F\bigl(\phi(t_s,x)\bigr)
  = \frac{\delta S[\phi]}{\delta\phi(t_s,x)}
  = -\partial^2\phi(t_s,x) + (m^2+\delta m^2)\phi(t_s,x)
    + \frac{g_0}{3!}\phi(t_s,x)^3.
\end{equation}
We have written the bare mass \(m_0\) in terms of the renormalized mass and its
counterterm (q.v.,~(\ref{eq:dm2})).  In the above equations \(t_s\) is the so
called stochastic (or simulation) time in which the stochastic field \(\phi\)
evolves. The field \(\eta\) is a field of Gaussian random numbers
satisfying\footnote{We use the same notation for the random field correlation
  functions as in ISPT. We believe that no confusion is possible as it should
  be clear from the context, as well as from the different indices, which field
  we are referring to.}
\begin{equation}
  \bigl\langle\eta(t_s,x)\bigr\rangle_{\eta}=0,\qquad 
  \bigl\langle\eta(t_s,x)\eta(t_s',y)\bigr\rangle_{\eta}
    = 2\delta(t_s-t_s')\delta_{xy}.
\end{equation}
Through the Langevin equation (\ref{eq:Langevin}) the field \(\phi\) 
depends upon the random field~\(\eta\).  The main assertion of stochastic
quantization is that the following identity holds order by order in
perturbation theory:
\begin{equation}
  \label{eq:ExpectationValueLangevin1}
  \lim_{t_s\to\infty} \bigl\langle\phi(t_s,x_1)\cdots
    \phi(t_s,x_n)\bigr\rangle_{\eta} 
    = \bigl\langle \varphi(x_1)\cdots\varphi(x_n)\bigr\rangle.
\end{equation}
Hence, in the long stochastic time limit the equal time correlation functions
of the stochastic field \(\phi\) converge to the expectation values
(\ref{eq:ExpectationValue}) of the Euclidean field theory with action \(S\); in
particular the equilibrium probability distribution of the stochastic field
\(\phi\) is proportional to~\(e^{-S(\phi)}\).  Equivalently, one can say that
in this limit the Langevin equation effectively trivializes the original theory
(q.v.,~(\ref{eq:Trivialization})).

Stochastic perturbation theory amounts to solving the Langevin equation
(\ref{eq:Langevin}) order by order in the couplings of the theory; in our case
these are \(g_0\) and \(\delta m^2\).  Substituting the expansion of the
stochastic field \(\phi\) analogous to (\ref{eq:DoublePowerSeriesField}) into
(\ref{eq:Langevin}) gives a system of equations for the fixed order fields,
\begin{equation}
  \begin{split}
    \partial_{t_s} \phi_{0,0}(t_s,x)
      &= (\partial^2 - m^2)\phi_{0,0}(t_s,x) +\eta(t_s,x), \\
    \partial_{t_s} \phi_{1,0}(t_s,x)
      &= (\partial^2 - m^2)\phi_{1,0}(t_s,x) -\frac1{3!}\phi_{0,0}(t_s,x)^3 ,\\
    \partial_{t_s} \phi_{0,1}(t_s,x)
      &= (\partial^2 - m^2)\phi_{0,1}(t_s,x) - \phi_{0,0}(t_s,x),
  \end{split}
\end{equation}
and so on. These equations can readily be solved for the \(\phi_{k,\ell}\)
fields. Once a solution is obtained up to a given order in the coupling,
(\ref{eq:ExpectationValueLangevin1}) can be used to compute the perturbative
expansion of any correlation function in the corresponding Euclidean field
theory (see~\cite{Damgaard:1987rr} for explicit examples of such calculations).

LSPT is the numerical implementation of this idea.  Stochastic time is
discretized as \(t_s = n\varepsilon\), with \(n\in\mathbb{N}\) and
\(\varepsilon\) being the step-size; a solution of the (discrete) Langevin
equation is then obtained according to some given integration scheme.  The
simplest such solution is provided by the Euler scheme, which is defined by the
update step
\begin{equation}
  \label{eq:EulerScheme}
  \phi\bigl((n+1)\varepsilon, x\bigr)
  = \phi(n\varepsilon,x) - \varepsilon F\bigl(\phi(n\varepsilon,x)\bigr)
    + \sqrt{\varepsilon}\,\eta(n\varepsilon,x),
\end{equation}
where here the random field \(\eta\) is normalized such that \(\bigl\langle
\eta(n\varepsilon,x) \eta(n'\varepsilon,y)\bigr\rangle_\eta = 2\delta_{nn'}
\delta_{xy}\), and \(\phi(0,x)\) is some given initial condition.  The
perturbative expansion of this solution is performed in an automated fashion by
employing order by order operations analogous
to~(\ref{eq:OrderByOrderProduct}); once this is given the expansion of a
generic observable \(\obs\bigl(\phi(t_s)\bigr)\) is obtained in the same way as
in~(\ref{eq:StochasticObservable}).  Assuming ergodicity the average over the
random field distribution in (\ref{eq:ExpectationValueLangevin1}) is replaced
by an average over stochastic time, and one obtains
\begin{equation}
  \label{eq:ExpectationValueLangevin2}
  \lim_{t_s\to\infty} \bigl\langle \obs(t_s)\bigr\rangle_\eta
    = \langle\obs\rangle
    \quad\xrightarrow{t_s = n\varepsilon}\quad
  \lim_{T\to\infty} \frac1T \sum_{n=0}^T \obs\bigl(\phi(n\varepsilon)\bigr)
    = \langle\obs\rangle + \order(\varepsilon^p).
\end{equation}
In the above relation the equivalence between correlation functions is valid
order by order in perturbation theory (q.v.,
(\ref{eq:PowerSeriesGreenFunction})), whereas the power \(p\) depends on the
order of the chosen integration scheme (see below).\footnote{In practical
  simulations the value of \(T\) is necessarily finite, and one averages the
  fields only once the discrete stochastic process has equilibrated.}

As asserted earlier, stochastic estimates of perturbative expansions of the
correlation functions of the target theory are obtained by use of Monte Carlo
sampling based on the Langevin equation.  We note that within the statistical
uncertainties the perturbative expansions so obtained are correct only up to
systematic errors due to the discretization of the stochastic time.  As
anticipated in (\ref{eq:ExpectationValueLangevin2}) these corrections are
expected to vanish as some power of the step-size as
\(\varepsilon\to0\)~\cite{Batrouni:1985jn,Kronfeld:1992jf}. The rate of
convergence depends on the choice of the numerical integrator employed for the
solution of the Langevin equation.  Such integrators are normally devised in
such a way that the discrete stochastic process associated with the given
integration scheme of order \(p\) converges, for small enough \(\varepsilon\),
to an equilibrium probability distribution \(\bar P(\phi) \propto e^{-\bar
  S(\phi)}\) where \(\bar S = S+\Delta S\) with \(\Delta S =
\order(\varepsilon^p)\).  Such deviation from the desired equilibrium
distribution is the cause of the corrections in the expectation value in
(\ref{eq:ExpectationValueLangevin2}) (see,
e.g.,~\cite{Kronfeld:1992jf,ZinnJustin:2002ru} for more details). In this work
we used a second order Runge--Kutta integrator (RK2): its exact definition is
given by eqs.~(A.4) and (A.15) of~\cite{Aarts:2011zn}.%
\footnote{We note that the RK2 integrator considered here requires \emph{three}
	  force computations per step.}
Using this integrator one expects corrections of \(\order(\varepsilon^2)\) in 
the perturbative computation of any correlation function.

It is clear that compared to ISPT the cost of LSPT with the perturbative order
in the couplings is rather mild. This is dictated by the order-by-order
operations necessary to integrate the discrete Langevin equation. Consequently,
the computational cost of LSPT increases (roughly) with the square of the order
in each coupling (q.v.,~(\ref{eq:OrderByOrderProduct})). However, as just
mentioned, the results need to be extrapolated to zero in the step-size to
eliminate systematic errors in the results. In addition, as the fields entering
in the average in (\ref{eq:ExpectationValueLangevin2}) are generated by a
Markov process, the successive field configurations are correlated; this
increases the statistical error for a fixed number of field configurations.
These correlations need to be properly taken into account in order to obtain
valid error estimates for the results.  Their magnitude is expected to grow
proportionally to \(L^2\) as the continuum limit of the theory is approached
(see, e.g.,~\cite{Kronfeld:1992jf,Baulieu:1999wz,Luscher:2011qa}).  This result
is valid for any perturbative order \(\obs_{k,\ell}\bigl(\phi_{0,0}(t_s),
\ldots, \phi_{k,\ell}(t_s)\bigr)\) of the generic (multiplicatively
renormalizable) stochastic field \(\obs\bigl(\phi(t_s)\bigr)\), and follows from
the remarkable property that the Langevin equation is
renormalizable~\cite{ZinnJustin:1986eq,ZinnJustin:1987ux}
(see~\cite{Luscher:2011qa} for a discussion). This feature allows one to infer
the scaling behavior of Langevin-based algorithms not only in the free case
where \(g_0=0\) but also in the full interacting theory.  In particular, as
recently shown by Martin L\"uscher~\cite{LuscherNotes:2015}, the
renormalizability of the Langevin equation also allows one to conclude that the
variances of these coefficients, \({\mathop{\rm Var}}(\obs_{k,l}) =
\lim_{t_s\to\infty} \left(\bigl\langle\obs_{k,\ell}^2(t_s)\bigr\rangle_\eta -
\bigl\langle\obs_{k,\ell}(t_s)\bigr\rangle_\eta^2\right)\), are at most
logarithmically divergent when taking the continuum limit.  This property is
quite remarkable and is not guaranteed for other NSPT implementations.



\section{NSPT based on GHMD algorithms} \label{sec:GHMD}

The idea of stochastic perturbation theory is not limited to the Langevin
equation.  Any stochastic differential equation (SDE) which satisfies an analogous
property to (\ref{eq:ExpectationValueLangevin1}) can provide a way of
performing stochastic perturbation theory.  One interesting example is given by
the stochastic molecular dynamics (SMD) equations (\ref{eq:Kramers}).  In this
context these were first considered in \cite{Horowitz:1985kd}, and were
recently studied in detail in \cite{Luscher:2011qa}.  Similarly, one can set-up
perturbation theory in terms of the Hybrid Molecular Dynamics (HMD)
equations~\cite{Luscher:2011qa}.  This observation suggests the possibility of
defining NSPT based on the discretization of these SDEs or of ergodic variances
of the molecular dynamics (MD) equations; such as the Kramers
\cite{Horowitz:1986dt,Horowitz:1991rr,Jansen:1995gz,Luscher:2011kk} and HMD
algorithm respectively~\cite{Duane:1987de}.  Experience with conventional
non-perturbative lattice field theory simulations would suggest the advantages
of reformulating NSPT in terms of these algorithms rather than Langevin-based
ones.  However determining their efficiency in this context, in particular
their continuum scaling, is not a trivial issue.  The results for the free
field theory~\cite{Kennedy:2000ju} provide a complete understanding of the
lowest perturbative order dynamics.  On the other hand the lack of
renormalizability of the SMD and HMD equations~\cite{Luscher:2011qa} in general
precludes analytic control over the continuum scaling of these algorithms in
the interacting theory.  In the case of NSPT this means a lack of control of
the behaviour of the higher-order fields.  Consequently, the efficiency of
these algorithms in the context of NSPT must be addressed numerically; in
particular the situation could be substantially different from both the free
case and the case where the full theory is simulated.

In this section we define NSPT in terms of the HMD and Kramers algorithms
(see~\cite{Kennedy:2000ju} and references therein for their definition).  These
are all inexact algorithms, as we do not know how to add a Metropolis step that
would be valid for arbitrary values of the coupling beyond leading (free field)
order.  We could consider the more general Generalized Hybrid Molecular
Dynamics algorithm~\cite{Kennedy:2000ju}, but based on both the expectations
from free field theory and from non-perturbative lattice field theory
simulations the HMD and Kramers algorithms appear to be natural sub-classes of
the GHMD algorithm to consider.  We shall assume the reader to be familiar with
these algorithms, and we limit ourselves to describing the required
modifications for their NSPT formulations.  These algorithms will be called
HSPT and KSPT, respectively.

\subsection{HSPT}
\label{subsec:HSPT}

In the case of the HMD algorithm, the basic field evolution is described by the
MD equations,
\begin{equation}
  \label{eq:MD}
  \partial_{t_s} \phi(t_s,x) = \pi(t_s,x),\qquad
  \partial_{t_s} \pi(t_s,x) = -F\bigl(\phi(t_s,x)\bigr),
\end{equation}
where \(F\bigl(\phi(t_s,x)\bigr)\) is given by (\ref{eq:Sder}), and \(\pi\) is
the momentum field conjugate to~\(\phi\).  Similarly to the Langevin case
(cf.~\S\ref{sec:LSPT}), in the context of NSPT both fields \(\phi\) and \(\pi\)
are assumed to have an expansion of the form~(\ref{eq:DoublePowerSeriesField}).
All operations in the following are thus intended to be performed in an order
by order fashion (q.v.,~(\ref{eq:OrderByOrderProduct})).

As is well known an algorithm based on the MD equations alone conserves
``energy" and so is not ergodic: the latter needs to be supplemented by an
occasional refreshment of the momentum field.  Therefore the momentum field
\(\pi\) is sampled from a Gaussian distribution with zero mean and unit
variance at the beginning of each trajectory (\(t_s=t_0\)); the refreshed
momentum initially only has a non-zero lowest order component.  In formulas
\begin{equation}
  \label{eq:refresh}
  \bigl\langle \pi_{0,0}(t_0,x)\bigr\rangle_\pi = 0,\qquad
  \bigl\langle \pi_{0,0}(t_0,x)\pi_{0,0}(t_0,y)\bigr\rangle_\pi = \delta_{xy}, 
\end{equation}
and \(\pi_{k,\ell}(t_0,x)=0\) if either \(k>0\) or \(\ell>0\), where
\(\langle\cdots\rangle_\pi\) denotes the average over the momentum field
distribution at the beginning of a trajectory.  The momentum field will acquire
higher-order components during the MD evolution (\ref{eq:MD}) from the time
\(t_0\) at which it was refreshed to time \(t_s = t_0 + \tau\), where \(\tau\)
is the trajectory length.  Numerically the MD evolution is determined by
discretizing the simulation time as \(t_s = n\delta t\), with
\(n\in\mathbb{N}\) and \(\delta t\) the step-size, and employing a suitable
integration scheme (see below).  Expectation values of generic observables are
then obtained similarly to (\ref{eq:ExpectationValueLangevin2}) by averaging
over sequences of trajectories.

For the numerical integration of the MD equations it is convenient to rely on
some reversible symplectic integration scheme, even though this is not necessary in
principle.\footnote{From here on we will refer to reversible symplectic integrators
simply as symplectic integrators.}
Symplectic integrators can systematically be improved, and
sophisticated symplectic integrators are readily available
(q.v.,~\cite{Kennedy:2012gk} for a discussion).  Moreover, once an efficient
symplectic integrator is found for a scalar theory, it can be extended to
non-Abelian theories in a straightforward manner.  For this work we used the
fourth order integrator defined by equations~(63) and (71)
of~\cite{Omelyan:2013}, which we refer to as the OMF4 integrator.%
\footnote{We note that the OMF4 integrator requires \emph{six} force computations
	  per step.}
Given this choice of integrator we expect \(\order(\delta t^4)\) errors in the 
results. More precisely, we expect in general that the equilibrium probability
distribution of fields generated through the HMD algorithm with some symplectic
integrator of order \(p\) is, for small enough step-size \(\delta t\), of the
form \(\bar P(\phi) \propto e^{-\bar S(\phi)}\), where \(\bar S = S + \Delta
S\) with \(\Delta S = \order(\delta t^p)\) (see~\cite{Clark:2007ffa} for more
details).  Consequently, since \(\Delta S\propto V\), one may argue that in order
to keep the step-size errors in the equilibrium distribution (approximately) constant
as the system size \(V\) is increased, one needs to keep the quantity \(y\equiv
V\delta t^p\) fixed.  It is clear that this is feasible only if efficient
high-order integrators are employed.\footnote{As mentioned before, we could
  include an accept/reject step in the HMD evolution of the lowest order field
  \(\phi_{0,0}\).  In this case the equilibrium probability distribution would
  be correct at this order.  Keeping the acceptance probability fixed in this
  case would then require \(x=V\delta t^{2p}\) to be fixed, which is
  a less stringent condition than keeping \(y\) fixed.  However, it is not
  clear what the step-size errors would be for the higher-order components of
  the field in this case.} 
We note that although keeping $y$ fixed would keep systematic errors in generic
correlators approximately constant as the system size is increased, this is 
probably an over-conservative condition if one is interested in (connected) 
correlation functions of local fields~\cite{Batrouni:1985jn,Kronfeld:1992jf}.

The HSPT algorithm described so far is not yet ergodic, the problem being that
the evolution of the lowest-order (free) field \(\phi_{0,0}\) is not
ergodic~\cite{Mackenzie:1989us, Kennedy:2000ju}; this in turn affects the
evolution of the higher-orders.  The solution to this problem is simple and is
to randomize the trajectory length \(\tau\)~\cite{Mackenzie:1989us}.  The
choice of distribution for the trajectories lengths may affect the
efficiency of the algorithm.  In our implementation we fixed the
step-size \(\delta t\), while choosing the number of steps \(n\) composing the
trajectory according to a binomial distribution with mean \(\langle n\rangle\).
This defines the average trajectory length to be~\(\langle\tau\rangle = 
\langle n\rangle\delta t\).

We conclude by pointing out that if one chooses \(\tau = \delta t\), i.e., the
trajectory consists of a single step, then the HMD algorithm effectively
integrates the Langevin equation (\ref{eq:Langevin})
(q.v.,~\cite{Clark:2007ffa} and below).  In other words, in this case the
algorithm just described can be interpreted as a particular integration scheme
for the Langevin equation.

\subsection{KSPT}
\label{subsec:KSPT}

Having defined HSPT in terms of the HMD algorithm, a second interesting
possibility to consider is NSPT based on the Kramers algorithm.  This algorithm was
proposed long ago in the context of field theory simulations by
Horowitz~\cite{Horowitz:1986dt,Horowitz:1991rr}, and recently reconsidered
in~\cite{Luscher:2011kk}.  In this case, the stochastic equations governing the
fields dynamics are given by the SMD equations,
\begin{equation}
  \label{eq:Kramers}
  \partial_{t_s} \phi(t_s,x) = \pi(t_s,x),\qquad
  \partial_{t_s} \pi(t_s,x) = -\gamma\pi(t_s,x) - F\bigl(\phi(t_s,x)\bigr)
    + \eta(t_s,x).
\end{equation}
Here \(F\bigl(\phi(t_s,x)\bigr)\) is still defined by (\ref{eq:Sder}), while
\(\eta(t_s,x)\) is a Gaussian random field satisfying
\begin{equation}
  \bigl\langle\eta(t_s,x)\bigr\rangle_\eta = 0,\qquad
  \bigl\langle\eta(t_s,x)\eta(t_s',y)\bigr\rangle_\eta
    = 2\gamma\delta(t_s-t_s')\delta_{xy},
\end{equation}
where \(\gamma>0\) is a free parameter (see below).  We observe that 
the (non-ergodic) MD equations (\ref{eq:MD}) are obtained when~\(\gamma =
0\) while, up to a rescaling of stochastic time, the Langevin equation
(\ref{eq:Langevin}) is obtained for \(\gamma \to \infty\)
(q.v.,~\cite{Luscher:2011qa}).

The implementation of Kramers algorithm is as follows.  Starting from some
arbitrary initial values for the fields \(\phi(0,x)\) and \(\pi(0,x)\), the MD
equations corresponding to (\ref{eq:Kramers}) with \(\gamma=0\) are integrated
from \(t_s=0\) to \(t'_s=\delta t\) through a single step of a given numerical
integration scheme.  The value of \(\delta t\) thus defines the step-size of
the integrator.  After this MD step, the effect of the \(\gamma\) term and the
coupling to the random field \(\eta\) is taken into account by \emph{partially}
refreshing the momentum field: the momentum field \(\pi(t'_s,x)\) is replaced
by
\begin{equation}
  \label{eq:PartialRefreshment}
  \pi'(t'_s,x) =  e^{-\gamma\delta t}\pi(t'_s,x)
    + \sqrt{1-e^{-2\gamma\delta t}}\,\eta(t'_s,x),
\end{equation}
where the noise field is here normalized such that \(\bigl\langle \eta(n\delta
t,x)\eta(n'\delta t,y)\bigr\rangle_\eta = \delta_{nn'}\delta_{xy}\).  These
elementary steps are then alternated, and expectation values of generic
observables of the field are obtained as in
(\ref{eq:ExpectationValueLangevin2}) by averaging over a long Monte Carlo
history, after they have reached equilibrium.  In a KSPT implementation, the
fields \(\phi\) and \(\pi\) are assumed to have an expansion of the form
(\ref{eq:DoublePowerSeriesField}), and just as in the Langevin case the random
field \(\eta\) only has a lowest-order component.  Hence, during the partial
refreshment (\ref{eq:PartialRefreshment}) only the lowest-order component of
the momentum field \(\pi_{0,0}\) is affected by the random field \(\eta\),
while the higher-order components are just rescaled by the factor
\(e^{-\gamma\delta t}\).  In the case where \(\gamma\to\infty\) (the Langevin
limit) the algorithm described is just the single step HSPT algorithm.

Having defined the algorithm, some comments are in order.  First of all, as
shown by Horowitz's analysis~\cite{Horowitz:1986dt}, the partial momentum
refreshment (\ref{eq:PartialRefreshment}) integrates exactly the corresponding
terms in (\ref{eq:Kramers}).  Similarly to the case of HSPT, the systematic
errors that one expects in expectation values of the fields
(\ref{eq:ExpectationValueLangevin2}) are given by the integration of the MD
equations in discrete steps; in particular analogous conclusions apply for the
order of the step-size errors in the equilibrium probability distribution
(q.v.,~\S\ref{subsec:HSPT}).  For the present work, we employed the very same
OMF4 integrator that we used for HSPT: we therefore expect \(\order(\delta
t^4)\) step-size errors.

Secondly, one might na{\"i}vely conclude from the free field theory analysis
of~\cite{Kennedy:2000ju} that the KSPT algorithm just defined is not of much
interest as it is not expected to perform better than HSPT, at least close to
the continuum limit.  However, one has to note that the conclusions 
in~\cite{Kennedy:2000ju} refer to the exact implementation of these
algorithms, i.e., when a Metropolis accept/reject step is included.  This is
what leads to the critical exponent for the \emph{cost} of the algorithms being
\(z=1\) for HMC but \(z=3/2\) for the exact Kramers algorithm (KMC).  However,
in the case of NSPT one is limited to \emph{inexact} algorithms, so the
computations have to be performed in a parameter regime where the effect of
step-size errors on expectation values are smaller than some specified statistical
accuracy, as otherwise some extrapolation in the step-size would be necessary.
In this regime, corresponding to the case where the Metropolis acceptance
probability would be close to one, the two algorithms have in fact comparable
performances~\cite{Kennedy:2000ju}.\footnote{It is worth pointing out that even
  in the exact case, the critical exponent for Kramers in the free case can be
  improved by using higher-order integrators for the MD equations.}

KSPT is also interesting due to the following property. As mentioned before,
the SMD equations (\ref{eq:Kramers}) approach the Langevin equation
(\ref{eq:Langevin}) in the limit \(\gamma\to\infty\).  In lattice field theory,
this limit can be taken simultaneously with the continuum limit if \(\gamma\)
is kept fixed in lattice units while \(a\to0\)~\cite{Luscher:2011qa}.  In this
limit the algorithm described above integrates the Langevin equation as the
continuum limit of the theory is approached.  Consequently, the considerations
on the continuum scaling of the LSPT algorithms discussed in \S\ref{sec:LSPT}
directly apply to KSPT at fixed \(\gamma\).  Although the scaling of these
algorithms is expected to be the same, in the case of KSPT the parameter
\(\gamma\) may be fixed to some finite value for which the algorithm may be
more efficient.  This will be addressed in detail in the next section.



\section{Numerical results} \label{sec:numres}

In this section we present the results of our numerical investigation of the
methods described in \S\ref{sec:ISPT}---\S\ref{sec:GHMD}.  Our aim is to
provide a comparison of the techniques in order to identify their principal
advantages and disadvantages.  In \S\ref{subsec:tests} we compare the
perturbative results for some specific quantities obtained with the different
algorithms, in order to confirm their correctness and viability.  Once these
are established, in \S\ref{subsec:scaling}---\S\ref{subsec:CostScaling} we
study the continuum scaling of the errors of these perturbative coefficients as
computed by the various methods.

\subsection{Testing the methods} \label{subsec:tests}

\begin{figure}[hpbt]
  \centering
  \includegraphics[width=0.675\textwidth]{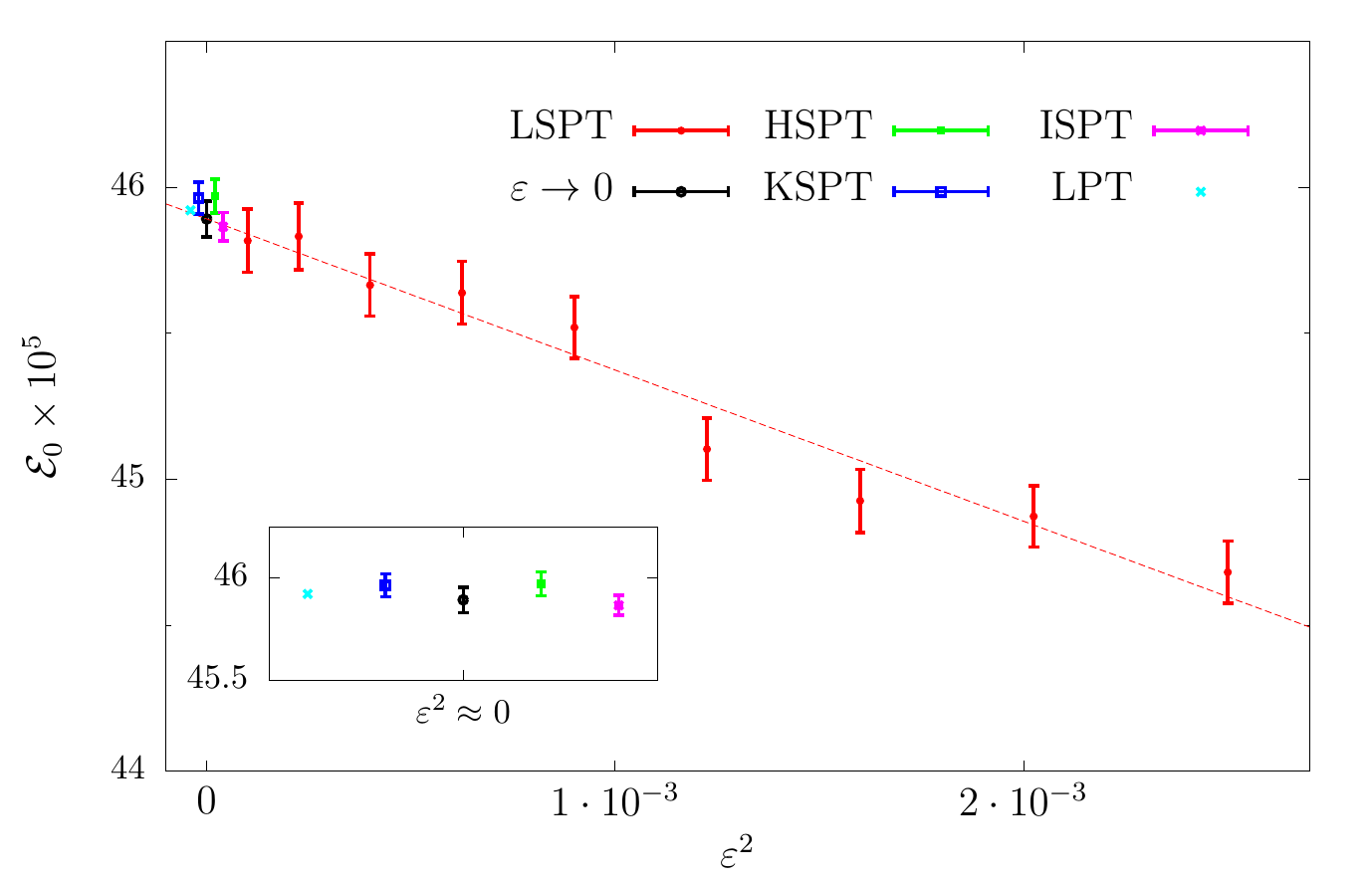}\\
  \includegraphics[width=0.675\textwidth]{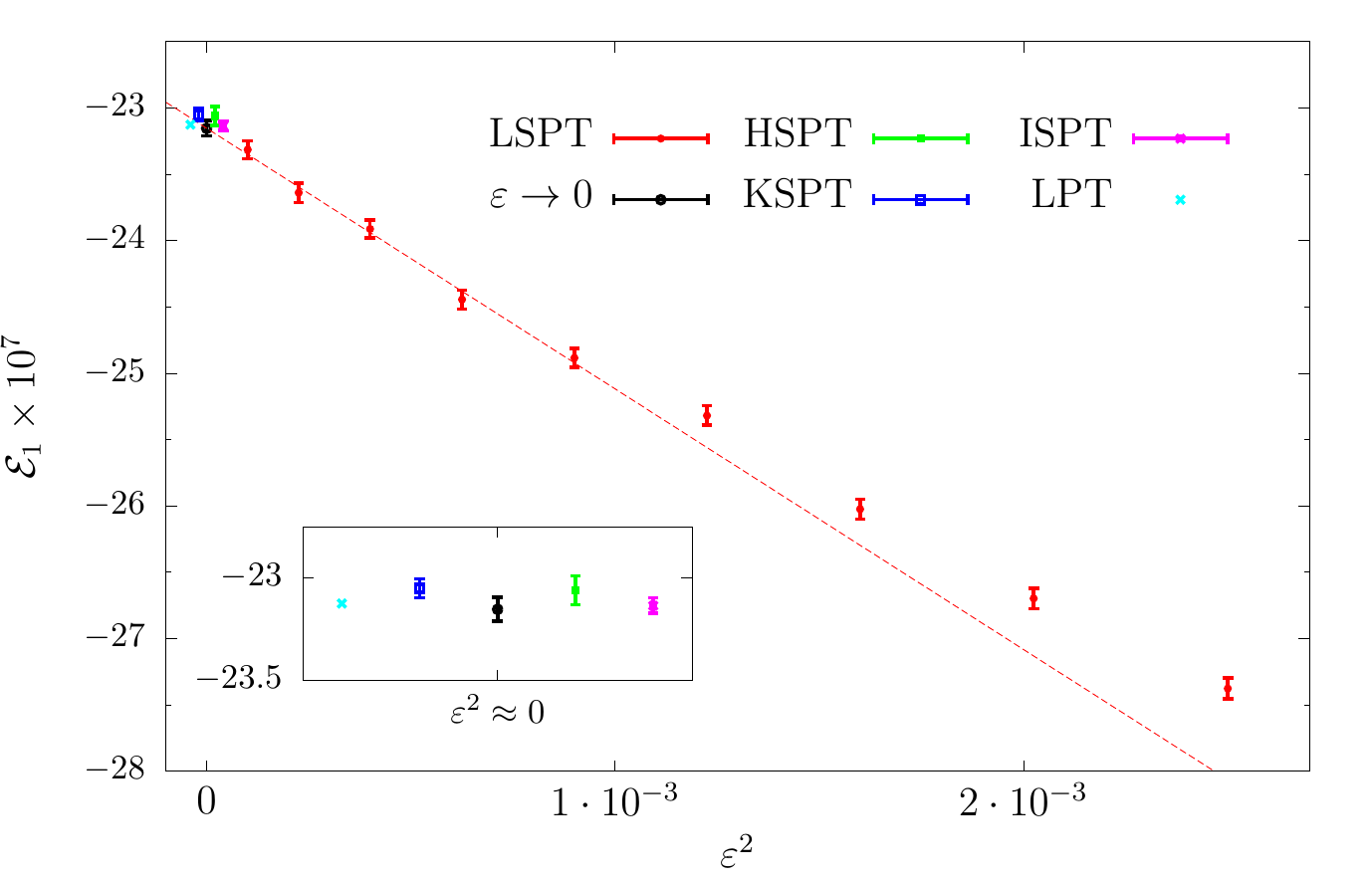}\\
  \includegraphics[width=0.675\textwidth]{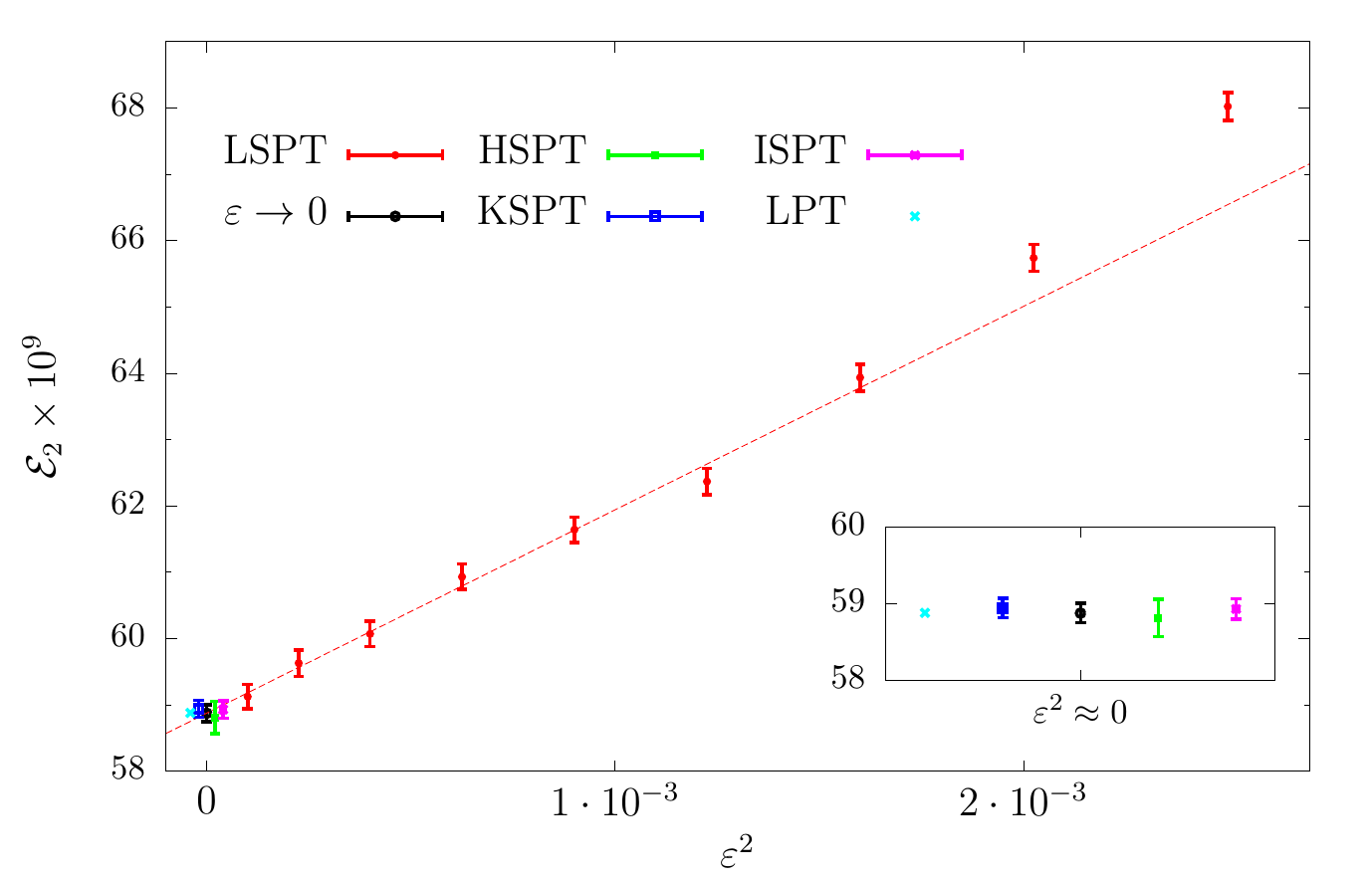}
  \caption{Comparison of different methods in the determination of
    \(\mathcal{E}_0\), \(\mathcal{E}_1\) and \(\mathcal{E}_2\) for \(z=4\),
    \(c=0.2\), and \(L=4\).  The analytic result (LPT) and the result of the
    extrapolation \(\varepsilon\to0\) for LSPT, as well as the ISPT, KSPT, and
    HSPT results (for which there are no step-size errors or the step-size
    errors are negligible compared with the statistical errors) are plotted
    near \(\varepsilon^2=0\).}
  \label{fig:Comparison}
\end{figure}

Before other comparisons are considered it is important to confirm that the
various algorithms agree for the perturbative computation of some quantities.
In Figure~\ref{fig:Comparison} the results for \(\mathcal{E}(t)\) at
tree-level, \(\order(g_0)\), and \(\order(g_0^2)\) are shown from top to bottom
respectively.  The computations were performed on a tiny \(L=4\) lattice for
which very high statistics could be obtained: similar results were obtained on 
larger lattices albeit with lower precision. We collected \(\approx 10^7\)
independent measurements for ISPT, HSPT, and KSPT, and \(\approx 10^6\)
measurements for each of 9 values of \(\varepsilon\in[0.01,0.05]\) for LSPT.
The values for the mass of the field and the flow time were chosen to
correspond to \(z=4\) and \(c=0.2\), respectively.  For HSPT and KSPT we then
chose \(\langle\tau\rangle=1\) and \(\gamma=2\).  The perturbative expansion is
expressed in terms of the renormalized mass \(m\) whereas the perturbative
coefficients correspond to the expansion in the bare coupling \(g_0\), i.e.,
\begin{equation}
  \label{eq:Ecal}
  \mathcal{E}(L,z,c) = \mathcal{E}_0 + \mathcal{E}_1g_0 + \mathcal{E}_2 g_0^2
    + \mathcal{E}_3g_0^3 + \order(g_0^4),\quad\mbox{where}\quad 
  \mathcal{E}_i \equiv \mathcal{E}_i(L,z,c).
\end{equation}
As can be seen from the figure, all the methods agree with each other and with
the analytic determination.  In the case of LSPT deviations from the expected
results are sizable at the largest step-sizes, and agreement is found only after
extrapolation to \(\varepsilon\to0\).  In particular, the asymptotic 
\(\order(\varepsilon^2)\) behaviour expected for the integrator used is clearly visible.

For the case of HSPT and KSPT we do not see any indication of step-size errors
as the results show no statistically significant deviation from the analytic
determination; the points are precise at the \(0.1\)--\(0.5\%\) level
depending on the order.  Even though the lattice is quite small, the step-size
we chose for both HSPT and KSPT is rather large, namely \(\delta t=0.5\).  This
step-size satisfies \(\delta t^4\geq 25\,\varepsilon^2\), for all values of
\(\varepsilon\) considered for LSPT: this inequality would give the na\"\i ve
size of the expected relative step-size errors.  This needs to be compared with
the fact that the application of the OMF4 integrator only costs twice as
much computer time as the RK2 integrator.
Of course this result depends on many factors: the lattice size considered, the 
observable, the parameters of the theory, the values of the step-sizes, and most
importantly the integrators used.\footnote{It is clear that considering larger 
lattices favours HSPT and KSPT, because higher-order integrators have a better 
cost scaling with increasing volume.} Nonetheless, as already emphasized, symplectic 
MD integrators are at a more mature stage of development than Runge Kutta integrators;
they can be optimized to reduce the \emph{magnitude} of the step-size errors 
(q.v.,~\cite{Omelyan:2013}).  As illustrated by our example, this results in a 
significant reduction of systematic errors relative to the cost of a single 
integration step.  Consequently, it is feasible to run the algorithm with a small
enough step-size such that extrapolations are not required. Moreover, as we can
afford to run with larger step-sizes for a fixed systematic error and with a fixed
number of force computations the cost of obtaining independent configurations is
reduced because of the smaller autocorrelations. Later in the section we shall give
more quantitative evidence on the benefits of using efficient symplectic integrators
in minimizing both systematic and statistical errors at fixed cost.

\subsection{Continuum error scaling: a first look} \label{subsec:scaling}

Having addressed the issue of systematic errors, we now study the continuum
scaling of the various NSPT algorithms.  We do this by investigating how the
(relative) errors of the perturbative coefficients of some given observables
scale as the continuum limit of the theory is approached.  The precise details
of the scaling depend on the observable, but some general features may be
inferred.

\begin{figure}[hptb]
  \centering
  \includegraphics[width=0.75\textwidth]{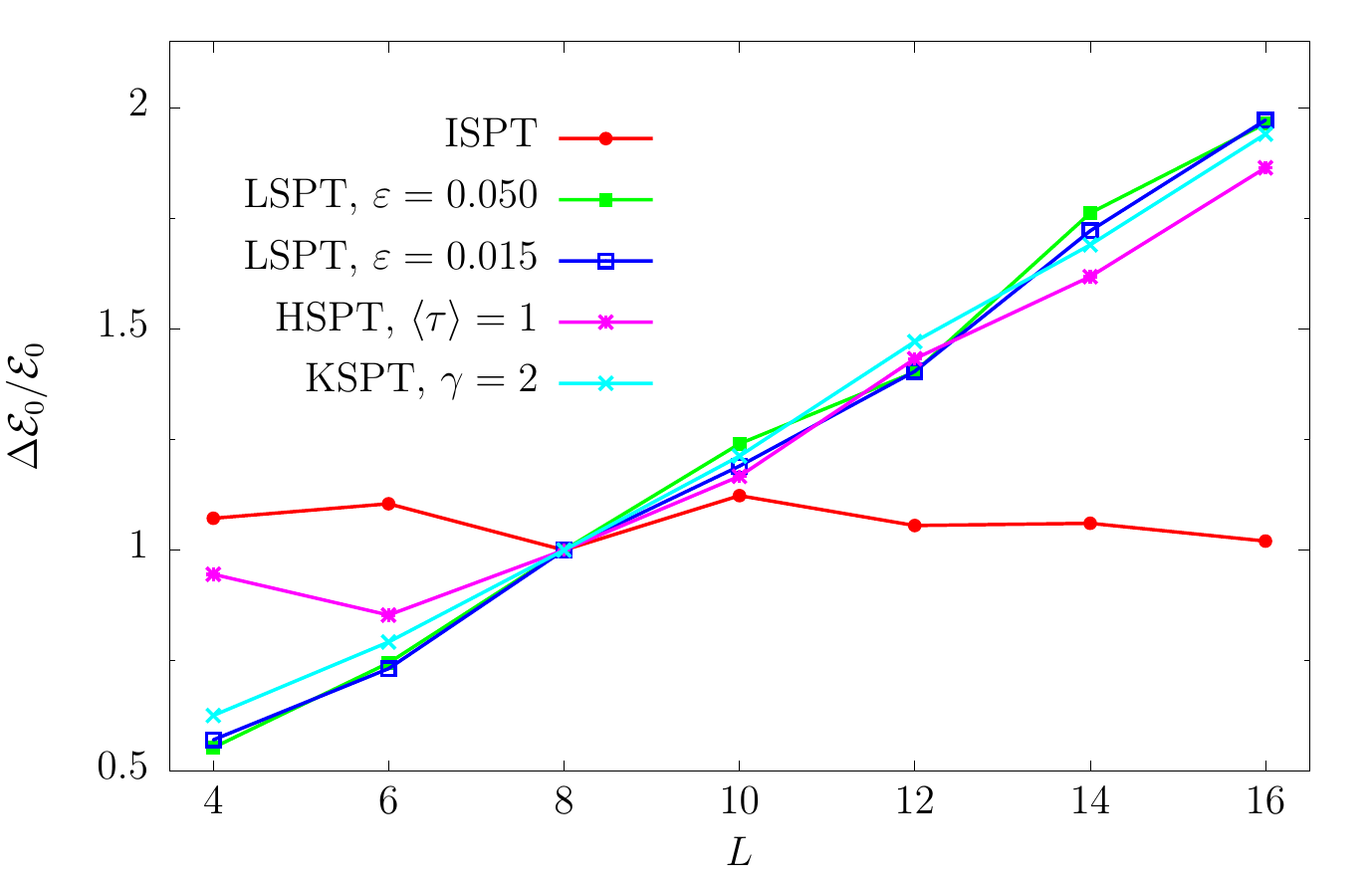}\\
  \includegraphics[width=0.75\textwidth]{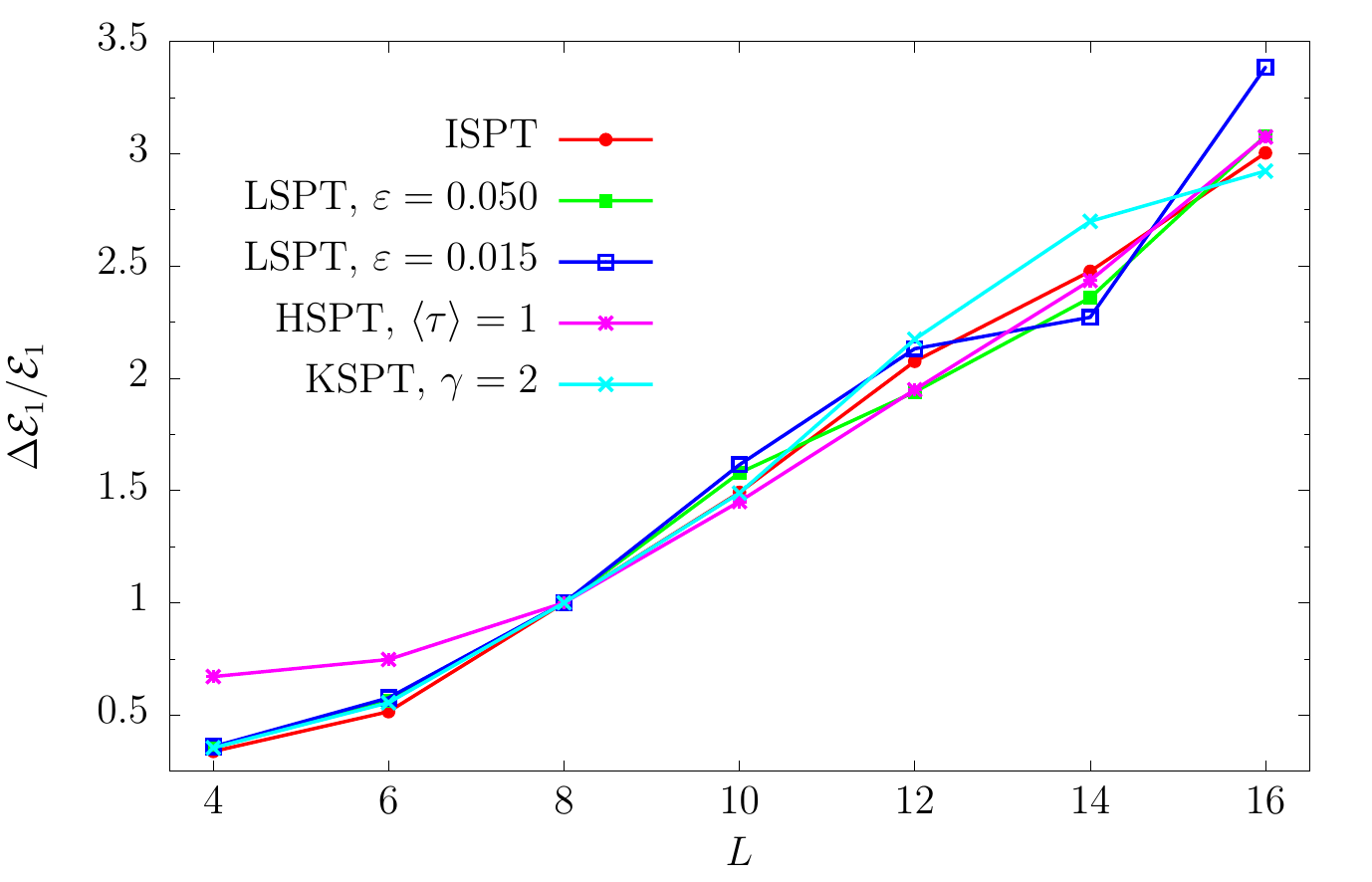}
  \caption{Continuum scaling of the relative errors \(\Delta \mathcal{E}_0 /
    \mathcal{E}_0\) and \(\Delta \mathcal{E}_1 / \mathcal{E}_1\) as computed
    with ISPT, LSPT, HSPT, and KSPT.  The parameters are \(z=4\) and \(c=0.2\).
    The data is normalized at \(L=8\).}
  \label{fig:RelativeError1}
\end{figure}

\begin{figure}[hptb]
  \centering
  \includegraphics[width=0.75\textwidth]{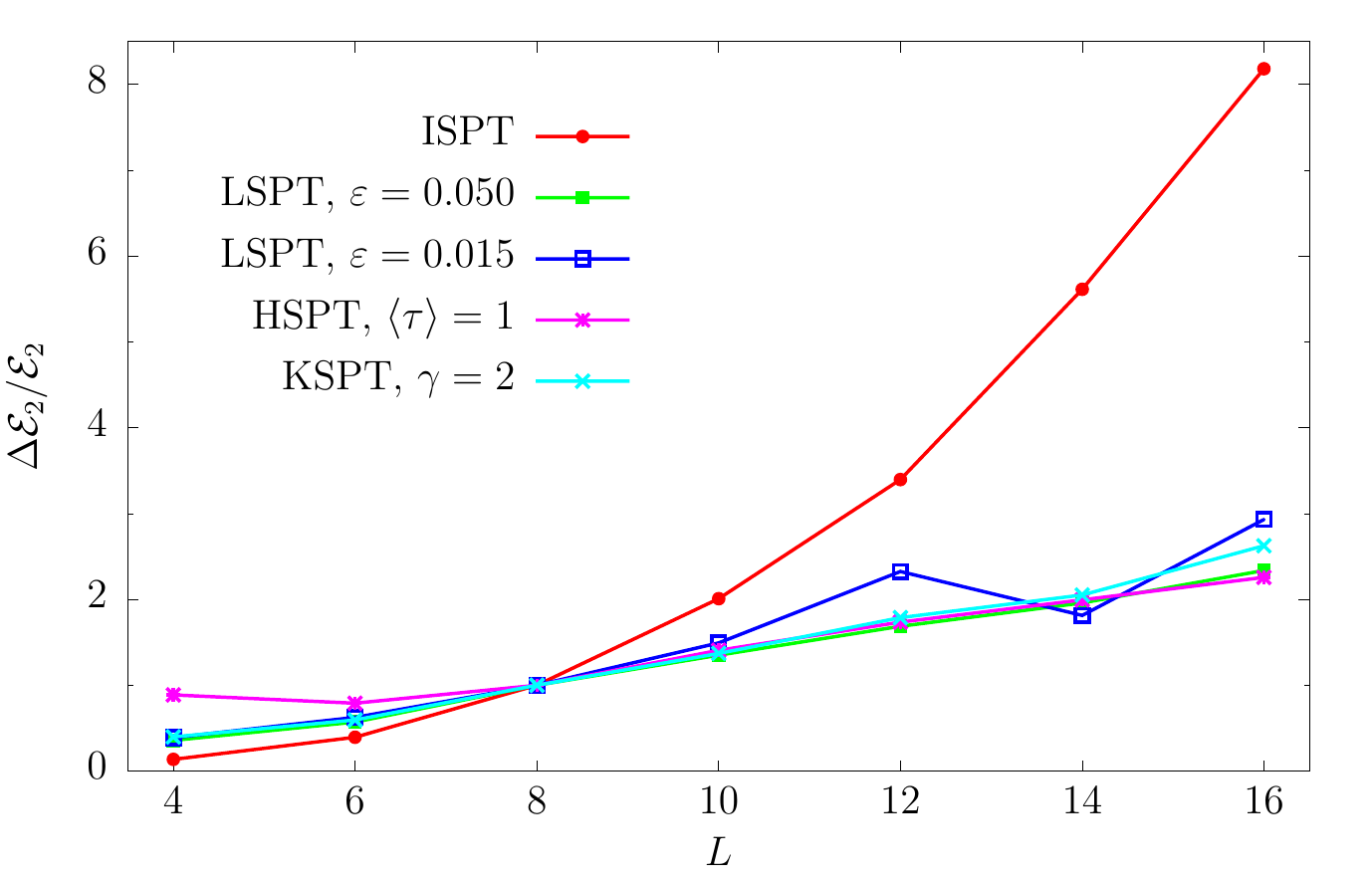}\\
  \includegraphics[width=0.75\textwidth]{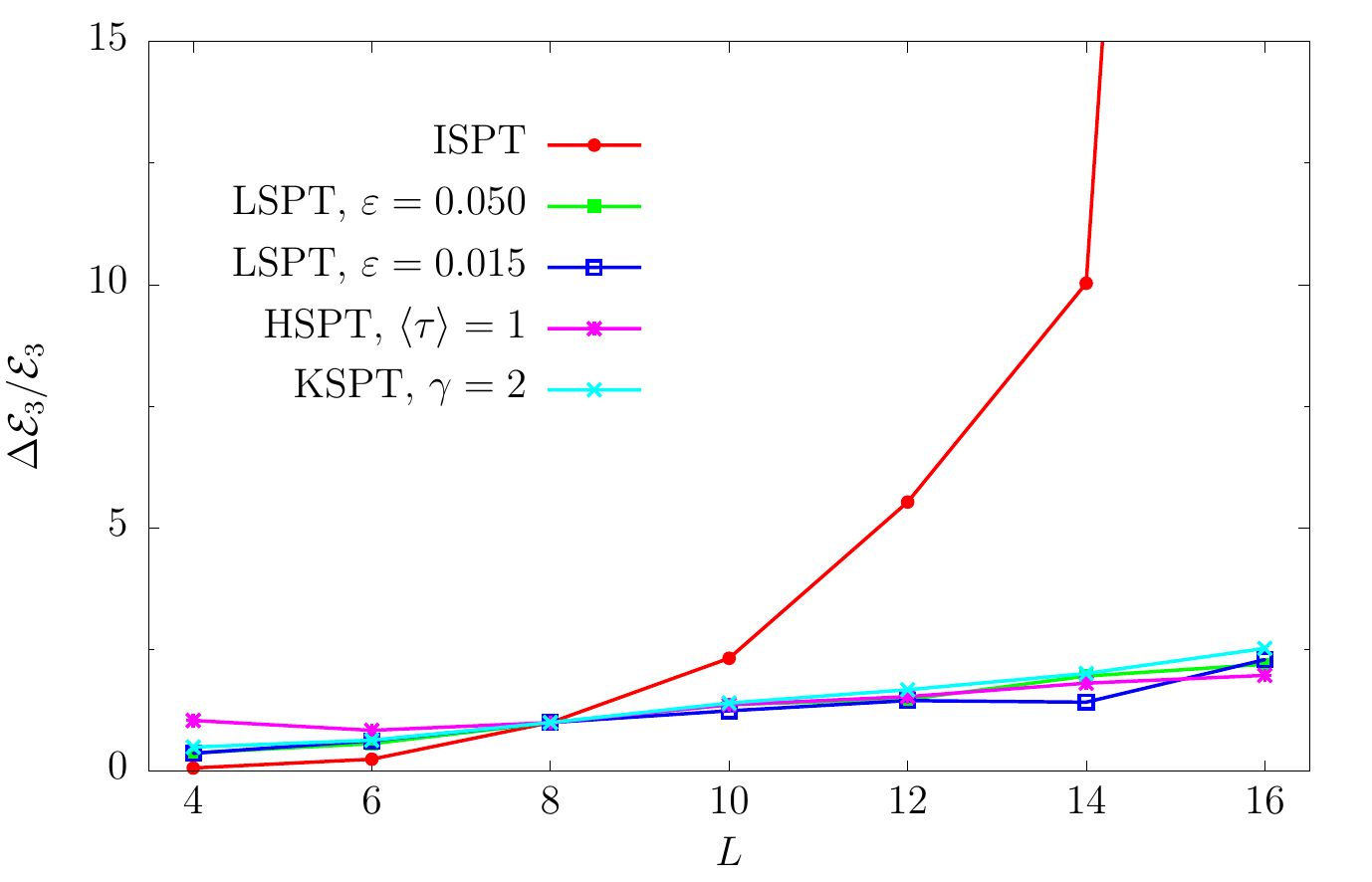}
  \caption{Continuum scaling of the relative errors \(\Delta \mathcal{E}_2 /
    \mathcal{E}_2\) and \(\Delta \mathcal{E}_3 / \mathcal{E}_3\) as computed
    with ISPT, LSPT, HSPT, and KSPT.  The parameters are \(z=4\) and \(c=0.2\).
    The data is normalized at \(L=8\).  Note that for ISPT \(\Delta
    \mathcal{E}_3/\mathcal{E}_3\approx65\) for \(L=16\).}
  \label{fig:RelativeError2}
\end{figure}

In Figures \ref{fig:RelativeError1} and \ref{fig:RelativeError2} we show the
continuum scaling of the relative errors \(\Delta \mathcal{E}_i /
\mathcal{E}_i\) for \(i=0,\ldots,3\) and \(L\) in the range \(4\leq L\leq16\),
as computed using ISPT, LSPT, HSPT, and KSPT.  Recall that the error \(\Delta
\mathcal{E}_i\) may be expressed as
\begin{equation}
  \label{eq:ErrorDefinition}
  \Delta \mathcal{E}_i = \sqrt{\frac{2A_I(E_i) \times
      \mathop{\rm Var}(E_i)} {N_{\mbox{\tiny config}}}},
\end{equation}
where \(N_{\mbox{\tiny config}}\) is the \emph{total} number of field
configurations considered, and \({\rm Var}(E_i)\) and \(A_I(E_i)\) are the
variance and integrated autocorrelation of \(E_i\) where
\begin{equation}
  \frac{t^2}{L^4} \sum_{x\in\Omega} E(t,x)
    = {E_0} + E_1 g_0 + E_2 g_0^2 + E_3g_0^3 +\order(g_0^4).
\end{equation}
We show our results for \(z=4\) and \(c=0.2\), but the same qualitative
behaviour is observed in other cases.  The number of configurations for each
method is specified at \(L=4\) and kept constant as \(1/L\to0\).  Specifically,
at \(L=4\) we collected between \(10^5\) and \(10^6\) independent measurements
for each of the different methods.  At this small lattice size and for the
algorithmic parameters considered the different methods have comparable
statistical precision for the same number of independent measurements.  In the
case of LSPT we measured after each step of the Markov chain.  For KSPT we set
the parameter \(\gamma=2\), and we adjusted the measurement frequency so as to
measure at fixed intervals \(\Delta t_s=0.5\) of simulation time independent of
the step-size.\footnote{Since autocorrelations are linear in the step-size
  \(\delta t\) for \(\gamma\) fixed, from the point of view of autocorrelations
  this is equivalent to measuring after each step for a fixed step-size of
  \(\delta t=0.5\).}  For HSPT we measured after each trajectory of average
length \(\langle\tau\rangle=1\).  The results in the figures are normalized to
the values of the relative errors at \(L=8\), and hence to a first
approximation are independent of~\(N_{\mbox{\tiny config}}\).  Since the
figures are only intended to be qualitative no estimates for the error on the
relative error are provided.

\begin{figure}[hptb]
  \centering
  \includegraphics[width=0.75\textwidth]{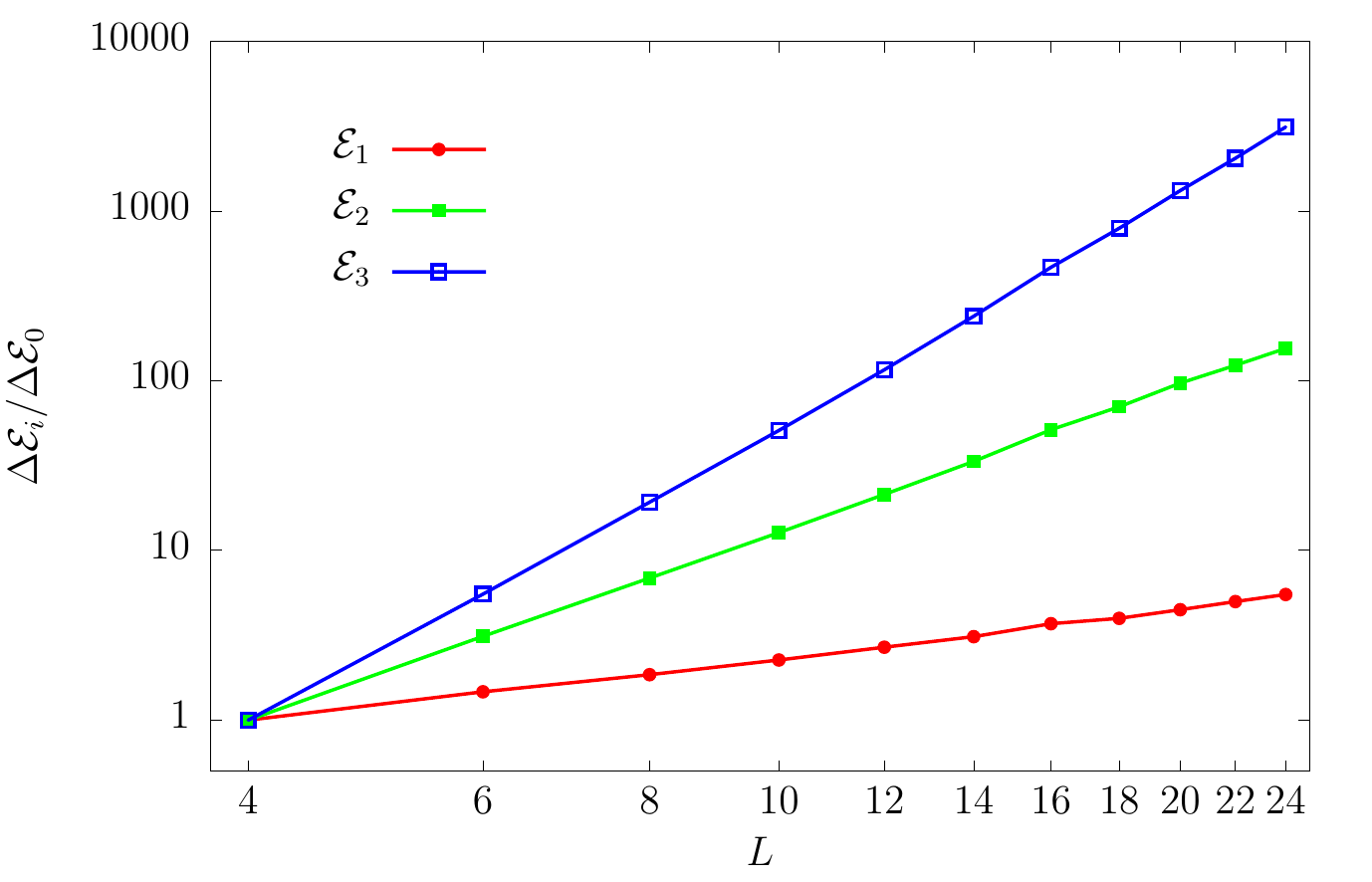}
  \caption{Continuum scaling of the ratios \(\Delta \mathcal{E}_i/\Delta
    \mathcal{E}_0\) for \(i=1,2,3\) as computed with ISPT.  The case with
    \(z=4\) and \(c=0.2\) is shown.  The results are normalized to their values
    at \(L=4\).}
  \label{fig:RelativeErrorISPT}
\end{figure}

The error computation for the perturbative coefficients was obtained using
jackknife in the case of ISPT, whereas for LSPT, HSPT, and KSPT we employed the
\(\Gamma\)-method described in~\cite{Wolff:2003sm} in order to take into
account autocorrelations of the measured quantities.  The coefficients
\(\mathcal{E}_i\) and corresponding errors refer to the expansion in terms of
the renormalized mass \(m\) and bare coupling \(g_0\)~(\ref{eq:Ecal}).  Power
divergences in the inverse lattice spacing are thus excluded in the
coefficients \(\mathcal{E}_i\), while logarithmic divergences associated with
renormalization of the coupling constant are not expected to be relevant for
the following discussion.  In the case of HSPT and KSPT the step-size was
scaled as \(\delta t \propto 1/L\) starting from a value of \(\delta t=0.5\) so
as to keep the \(\order(\delta t^4)\) errors in the equilibrium distribution
approximately fixed using the OMF4 integrator as the continuum limit is
approached. As mentioned in \S\ref{subsec:HSPT} this is probably a very conservative
choice, but it was done to avoid potentially large systematic errors that might 
modify the overall picture.\footnote{In fact with this choice the step-size errors
vanish faster than the leading \(O(1/L^2)\) lattice artifacts as the continuum limit
is approached.} Keeping the systematic errors in the equilibrium distribution fixed
for LSPT is significantly more challenging as it requires \(\varepsilon\propto 1/L^2\)
with the RK2 integrator (q.v., \S\ref{sec:LSPT}). In this case we thus simply considered
two well-separated step-sizes in order to assess the dependence of the results on~\(\varepsilon\).

Starting from the results at tree-level (top panel in Figure
\ref{fig:RelativeError1}) we see how the relative error of ISPT is constant for
a fixed number of field configurations.  The results for LSPT, HSPT, and KSPT
are rather different: excluding perhaps the smaller lattices there is a linear
growth of the relative errors with the lattice size.  These results confirm
free field theory expectations.  The variance \(\mathop{\rm Var}(E_0)\) is
finite and constant with \(L\) up to discretization effects.  In particular it
is the same for all NSPT methods up to step-size errors, and independent of the
algorithmic parameters.  Consequently, since ISPT results are uncorrelated,
this implies that the error \(\Delta \mathcal{E}_0\) is essentially constant
with \(L\) for a given number of field configurations \(N_{\mbox{\tiny
    config}}\).  The linear rise of the errors in the case of LSPT, HSPT, and
KSPT is due to the fact that autocorrelations grow \(\propto L^2\) as the
continuum limit is approached.  For a fixed number of configurations this
translates into a linear rise of the relative errors with \(L\) as the number
of independent configurations decreases \(\propto 1/L^2\).

At higher perturbative orders the situation for ISPT changes significantly.  At
\(\order(g_0)\) the relative error grows linearly with~\(L\), indicating a
growth of the variance proportional to \(L^2\) as the continuum limit is
approached.  For higher perturbative orders the increase of the variance is
even more rapid.  This may be better appreciated from
Figure~\ref{fig:RelativeErrorISPT} where results for ISPT alone are given up to
\(L=24\) and~\(\order(g_0^3)\).  In this plot we show the ratios of
\(\Delta\mathcal{E}_i\) and \(\Delta\mathcal{E}_0\) for \(i=1,2\) and \(3\).
These ratios are independent on the number of configurations considered, and were
estimated using \(10^5\)--\(10^7\) measurements, depending on the lattice size. 
It is clear that the error, and hence the variance, increases as an increasing power
of \(L\) as the perturbative order is increased.\footnote{A similar behaviour
  was also observed by Martin L\"uscher in pure SU(3) Yang-Mills
  theory~\cite{LuscherTalk:2015}.}  This is to be compared with the relative
errors for LSPT, HSPT, and KSPT, which have the same qualitative behaviour as
at tree level, namely the errors increase only linearly with~\(L\)
(Figures~\ref{fig:RelativeError1} and \ref{fig:RelativeError2}).  For these
algorithms the behaviour is similar to what happens at tree level: the errors
of the higher order coefficients appear to increase due to increasing
autocorrelations.  The increase of the variance of the perturbative
coefficients in LSPT, HSPT, and KSPT, if any, is very mild here.\footnote{We
  note that for LSPT a similar observation was made in~\cite{Brida:2013mva} in
  the pure SU(3) Yang-Mills theory.} These conclusions will be confirmed by the
detailed investigations of the following subsections.

We conclude by noticing that the above observations for the higher order
results are in agreement with general theoretical expectations.  The peculiar
behaviour in the variance of perturbative coefficients computed with ISPT was
recently elucidated by Martin L\"uscher~\cite{LuscherTalk:2015}.  He emphasized
the generic presence of power divergences in the variance of perturbative
coefficients computed with ISPT.  On the other hand, as we noted in
\S\ref{sec:LSPT}, he showed that the variances of perturbative coefficients
computed using LSPT are at most logarithmically divergent~\cite{LuscherNotes:2015}.  
However, their autocorrelations grow with the square of the correlation length
of the system, i.e., \(\propto1/m^{2}\propto L^2\).  These results also apply to
KSPT at fixed \(\gamma\) (q.v.,~\S\ref{subsec:KSPT} and~\cite{Luscher:2011qa}).
Strictly speaking they cannot be extended to HSPT due to the non-renormalizability
of the HMD equations~\cite{Luscher:2011qa}, although it is most plausible that
they hold in the case where the trajectory length does not scale with the
correlation length of the system, i.e., \(\langle\tau\rangle\) is independent
of \(L\).  This follows from the observation that in the continuum limit
\(L\to\infty\) the HSPT algorithm effectively integrates the perturbatively
expanded Langevin equation, as in this case there is no fundamental difference
from a single step HSPT algorithm (which is LSPT).  This conjecture seems to be
confirmed by the numerical experiments discussed below.

\subsection{Continuum scaling of autocorrelations}
\label{subsec:AutocorrelationScaling}

As a result of the investigation of the previous subsection we conclude that
NSPT methods based on stochastic differential equations have a much better
continuum cost scaling than ISPT.  It is clear that beyond the first few orders
in perturbation theory the scaling of ISPT is such that its performance is much
worse than the other algorithms.  In this and the following subsection we
therefore focus our attention on these other methods.  In particular the
question we want to address is the following.  As is well known, free field
analysis of the HMD and Kramers algorithms shows that their continuum cost
scaling depends on how their parameters are adjusted~\cite{Kennedy:2000ju}.  In
the context of NSPT these results directly apply to the lowest order
determinations.  However, it is not obvious what the behaviour of the higher
order results is if different parameter scalings are considered; this is
because we do not have analytic control on this behaviour except in the
Langevin limit of these algorithms.  To answer this question we investigate the
continuum scaling of the autocorrelations of the perturbative orders \({E_i}\)
as a function of the algorithmic parameters in this subsection.  More
precisely, we will compare the optimal parameter scaling suggested by the free
field theory analysis of~\cite{Kennedy:2000ju} with the Langevin scaling.  We
identify the latter as the case where \(\langle\tau\rangle\) for HSPT or
\(\gamma\) for KSPT is kept fixed as the continuum limit is approached.  The
case of LSPT is not considered explicitly as it is effectively covered by KSPT
for \(\gamma\to\infty\) or equivalently by a single-step HSPT algorithm.

Starting with HSPT, at the lowest perturbative order we expect autocorrelations
to grow like \(L^2\) when approaching the continuum limit if the average
trajectory length \(\langle\tau\rangle\) is kept fixed.  On the other hand, the
analysis of~\cite{Kennedy:2000ju} shows how this scaling can be improved by
choosing the average trajectory length proportional to the correlation length
of the system: \(\langle\tau\rangle\propto 1/m\propto L\).  Heuristically, the
idea is that by adjusting the trajectory length with the correlation length one
avoids the situation where configuration space is explored by a random walk,
namely in random steps that are short compared with the natural scale of the
system.  What happens to the autocorrelations in HSPT beyond the tree-level
dynamics, however, remains to be seen.\footnote{In the full theory this may not
  be the case~\cite{Luscher:2011kk}.}

\begin{figure}[hptb]
  \centering
  \includegraphics[width=0.75\textwidth]{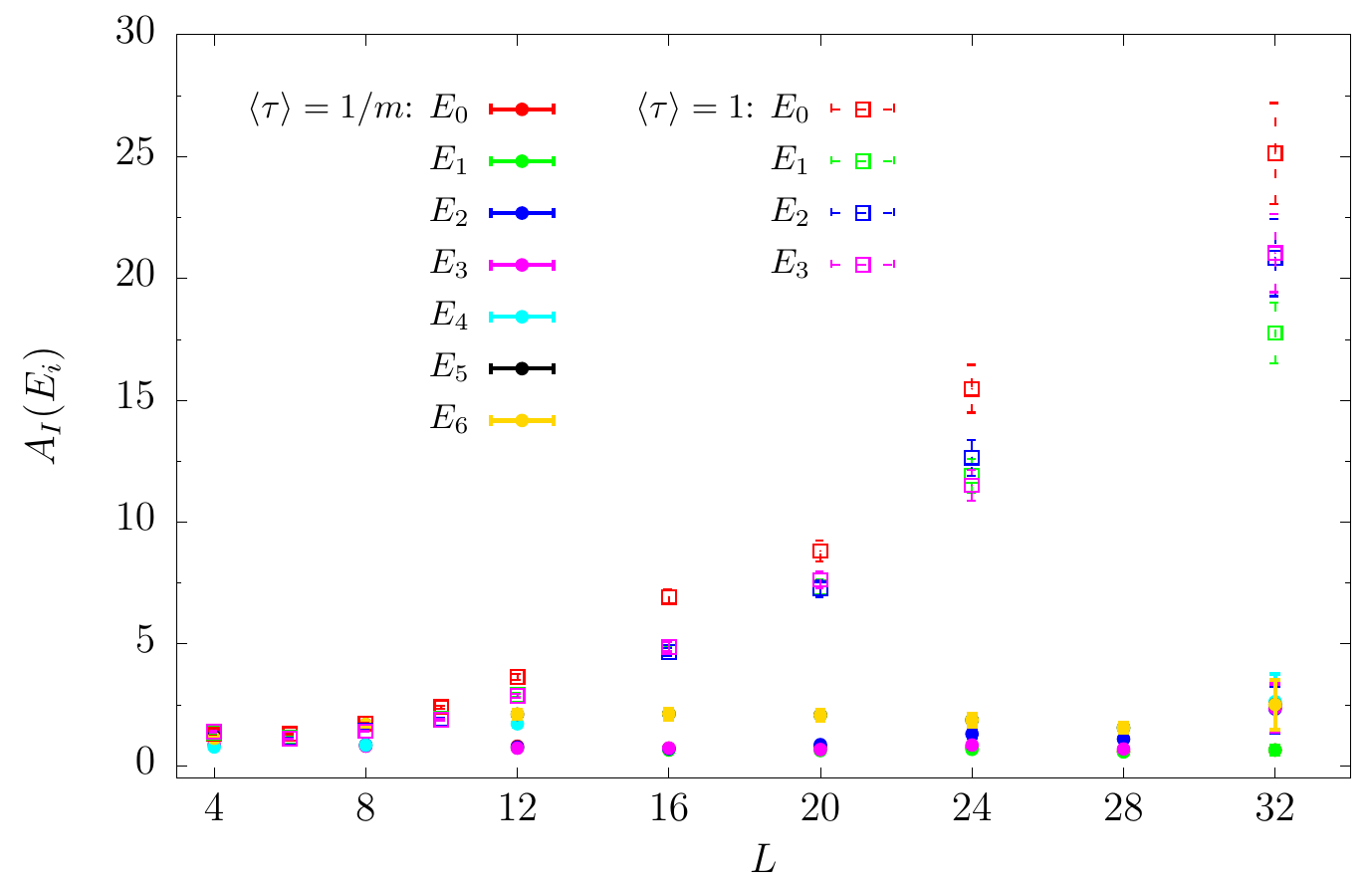}
  \caption{Continuum scaling of the integrated autocorrelations \(A_I(E_i)\) in
    HSPT for the cases \(\langle\tau \rangle=1\) and \(\langle\tau
    \rangle=1/m\).  For \(\langle\tau\rangle=1\) we show results only up to
    \(\order(g_0^3)\), while for \(\langle\tau\rangle=1/m\) they go up to
    \(\order(g_0^6)\).  The data is for \(z=4\) and \(c=0.2\). We measure the 
    observables after each trajectory. The errors on the integrated autocorrelations
    were estimated using the \(\Gamma\)-method~\cite{Wolff:2003sm}.}
  \label{fig:ScalingAutocorrelationsHMD}
\end{figure}

In Figure~\ref{fig:ScalingAutocorrelationsHMD} we compare the results for the
integrated autocorrelation \(A_I(E_i)\) of the perturbative orders \(E_i\) as
the continuum limit is approached.  We compared the case where the average
trajectory length was kept fixed at \(\langle\tau\rangle=1\) with the case
where we set \(\langle\tau\rangle=1/m\) for the range of lattice sizes \(4\leq
L \leq 32\).  The step-size was adjusted so as to keep the errors in the
equilibrium distribution roughly constant as \(L\) was increased, namely
\(\delta t=2/L\) using the OMF4 integrator.  We measured the observables after
each trajectory, and chose \(z=4\) and \(c=0.2\).  As can be seen from the
figure the free field theory expectation also applies for the high-order
fields: for the case where \(\langle\tau\rangle=1\) we observed the asymptotic
random walk behaviour \(A_I(E_i)\propto L^2\) whereas for \(\langle\tau\rangle
= 1/m\) the integrated autocorrelations were constant as the continuum limit
was approached.

\begin{figure}[hptb]
  \centering
  \includegraphics[width=0.75\textwidth]{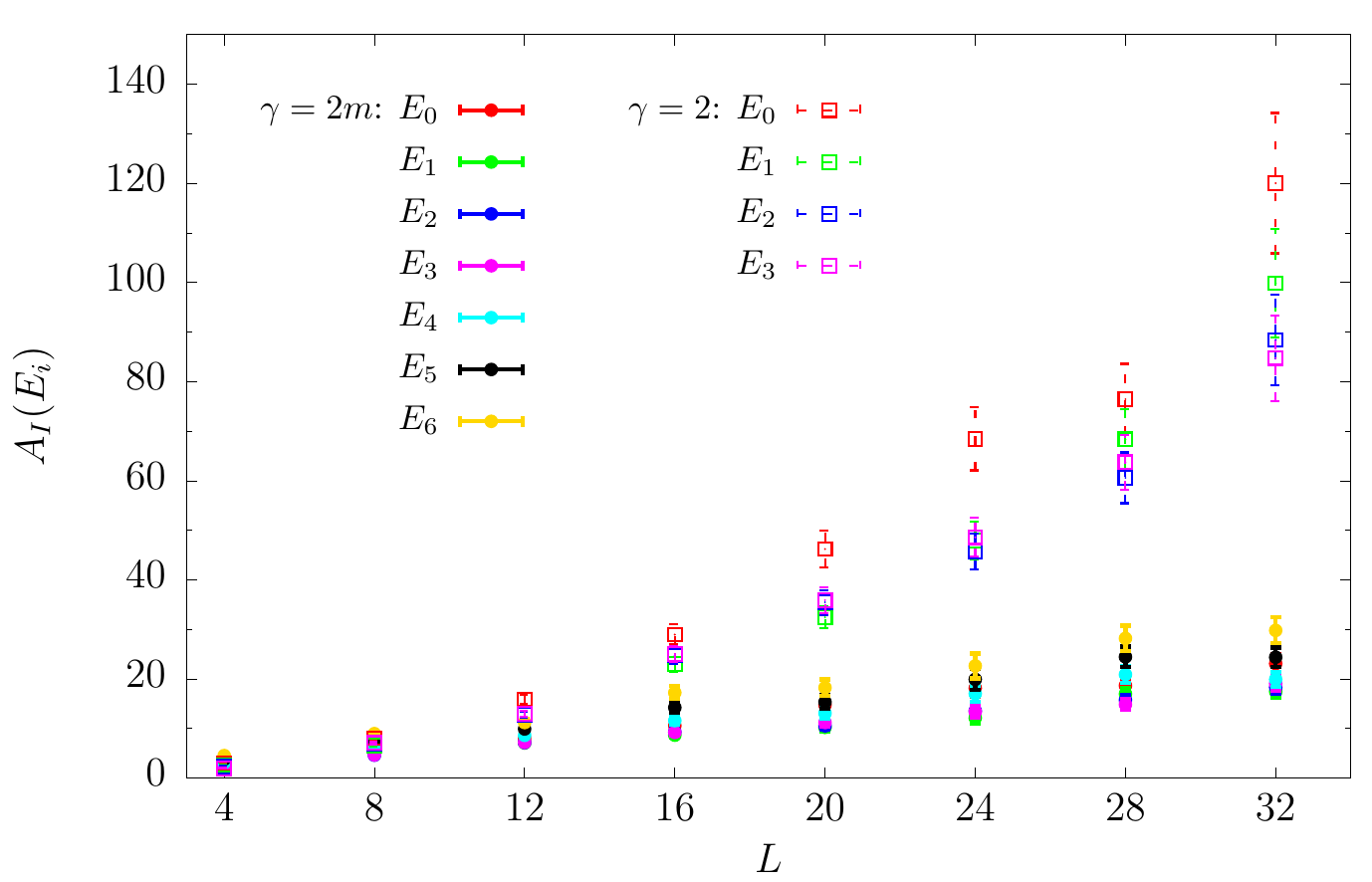}
  \caption{Continuum scaling of the integrated autocorrelations \(A_I(E_i)\) in
    KSPT for the cases \(\gamma=2\) and \(\gamma=2m\).  For \(\gamma=2\) we
    show results only up to \(\order(g_0^3)\), while for \(\gamma=2m\) they go
    up to \(\order(g_0^6)\).  The data is for \(z=4\), \(c=0.2\). The step-size 
    is \(\delta t=0.25\) and we measure the observables after each step.  The 
    errors on the integrated autocorrelations were estimated using the 
    \(\Gamma\)-method~\cite{Wolff:2003sm}.}
\label{fig:ScalingAutocorrelationsSMD}
\end{figure}

For KSPT the results from free field theory~\cite{Kennedy:2000ju} indicate that
at the lowest order in perturbation theory the autocorrelations are expected to
increase as \(L^2\) as the continuum limit is approached if the parameter
\(\gamma\) is kept fixed.  However, they increase only as \(L\) if
\(\gamma\propto m\) (see also~\cite{Luscher:2011kk}).\footnote{We assume that
  the observables are measured at fixed stochastic time intervals
  as~\(L\to\infty\).}  Hence \(\gamma\) effectively plays the role of an
inverse trajectory length for the algorithm~\cite{Horowitz:1986dt}.  In Figure
\ref{fig:ScalingAutocorrelationsSMD} we report the results for \(A_I(E_i)\) for
these two cases.  In the first case we fixed \(\gamma=2\) as~\(L\to\infty\),
while in the second case we set \(\gamma=2m\). We measured the observables after
each step, and chose \(z=4\) and \(c=0.2\). Unlike the case of HSPT we chose
a fixed step-size \(\delta t=0.25\), and we kept this constant as 
\(L\to\infty\).\footnote{We checked up to \(L=20\) that compatible
  results for the integrated autocorrelations were obtained if \(\delta
  t\propto 1/L\) and the autocorrelations measured in units of this step-size
  were rescaled~\(\propto L\).} 
As we can see from the figure the two cases agree with the free field theory
expectations for all the perturbative orders we investigated.

In conclusion, it seems that the free field theory expectations for
autocorrelations of the HMD and Kramers algorithms apply up to relatively high
perturbative orders in the corresponding NSPT implementations.\footnote{We also
  studied the dependence of the integrated autocorrelations \(A_I(E_i)\) on the
  step-size \(\delta t\) and \(\gamma\) for KSPT, and on \(\langle
  \tau\rangle\) for HSPT at fixed \(L\) and~\(m\).  In this case the free field
  theory predictions of~\cite{Kennedy:2000ju} also hold for all the
  perturbative orders we investigated.}  Except for the case of KSPT at fixed
\(\gamma\) this is a non-trivial result in view of the non-renormalizability of
the HMD and SMD equations~\cite{Luscher:2011qa,Luscher:2011kk}.

\subsection{Continuum variance scaling} \label{subsec:VarianceScaling}

\begin{figure}[hptb]
  \centering
  \includegraphics[width=0.75\textwidth]{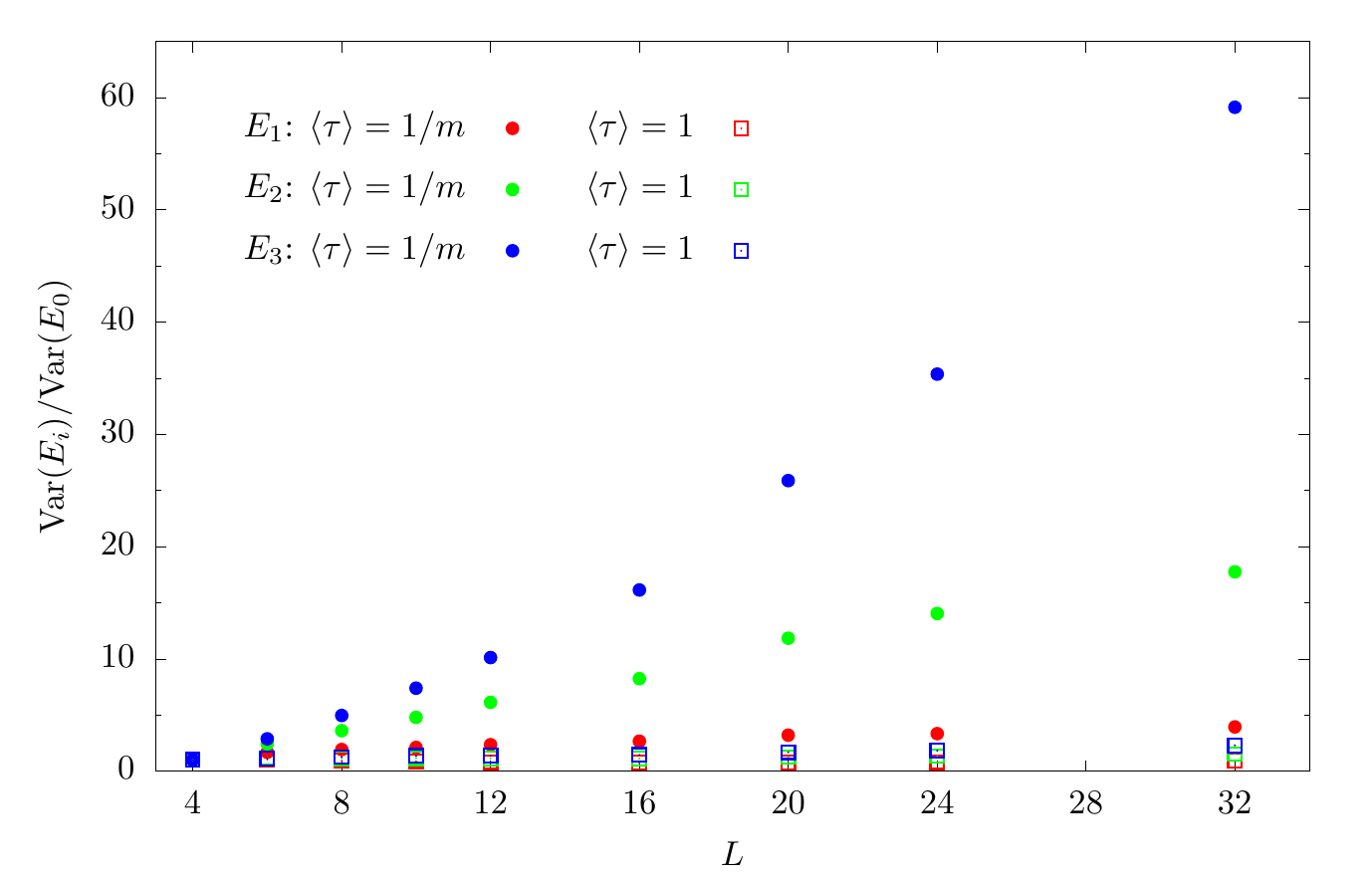}
  \caption{Continuum scaling of the ratios \(\mathop{\rm Var}(E_i) /
    \mathop{\rm Var}(E_0)\) with \(i=1,2,3\) for HSPT, for the cases
    \(\langle\tau\rangle=1/m\) and \(\langle\tau\rangle=1\).  The case of
    \(z=4\) and \(c=0.2\) is shown, and the data are normalized at \(L=4\).}
  \label{fig:ScalingVarianceHMD}
\end{figure}

Having investigated the dependence of the continuum scaling of the integrated
autocorrelations for different algorithmic parameter scalings, we next studied
the corresponding scaling of the variances \(\mathop{\rm Var}(E_i)\).  In
Figure~\ref{fig:ScalingVarianceHMD} we present results for the ratios
\(\mathop{\rm Var}(E_i)/\mathop{\rm Var}(E_0)\) with \(i=1,2,3\) for HSPT,
comparing the cases \(\langle\tau\rangle=1\) and \(\langle\tau\rangle=1/m\)
as~\(L\to\infty\).  For convenience the results are normalized by their values
at \(L=4\).  As usual we chose \(z=4\), \(c=0.2\), and took \(4\leq L\leq32\)
and~\(\delta t=2/L\).  Recall that the lowest order variance \(\mathop{\rm
  Var}(E_0)\) is independent of the algorithmic parameters, namely
\(\langle\tau\rangle\) (or \(\gamma\) below), and up to \(\order(1/L^2)\)
corrections is constant with~\(L\).  Observe that upon setting
\(\langle\tau\rangle=1/m\) the variances \(\mathop{\rm Var}(E_i)\) with \(i>1\)
increase significantly as the continuum limit is approached.  This effect is
more pronounced as the perturbative order increases; on the other hand for
\(\langle\tau\rangle=1\) the variances for all the perturbative orders
considered grow very slowly with \(L\) and do not change significantly over the
whole range of lattice sizes studied.

\begin{figure}[!hptb]
  \centering
  \includegraphics[width=0.75\textwidth]{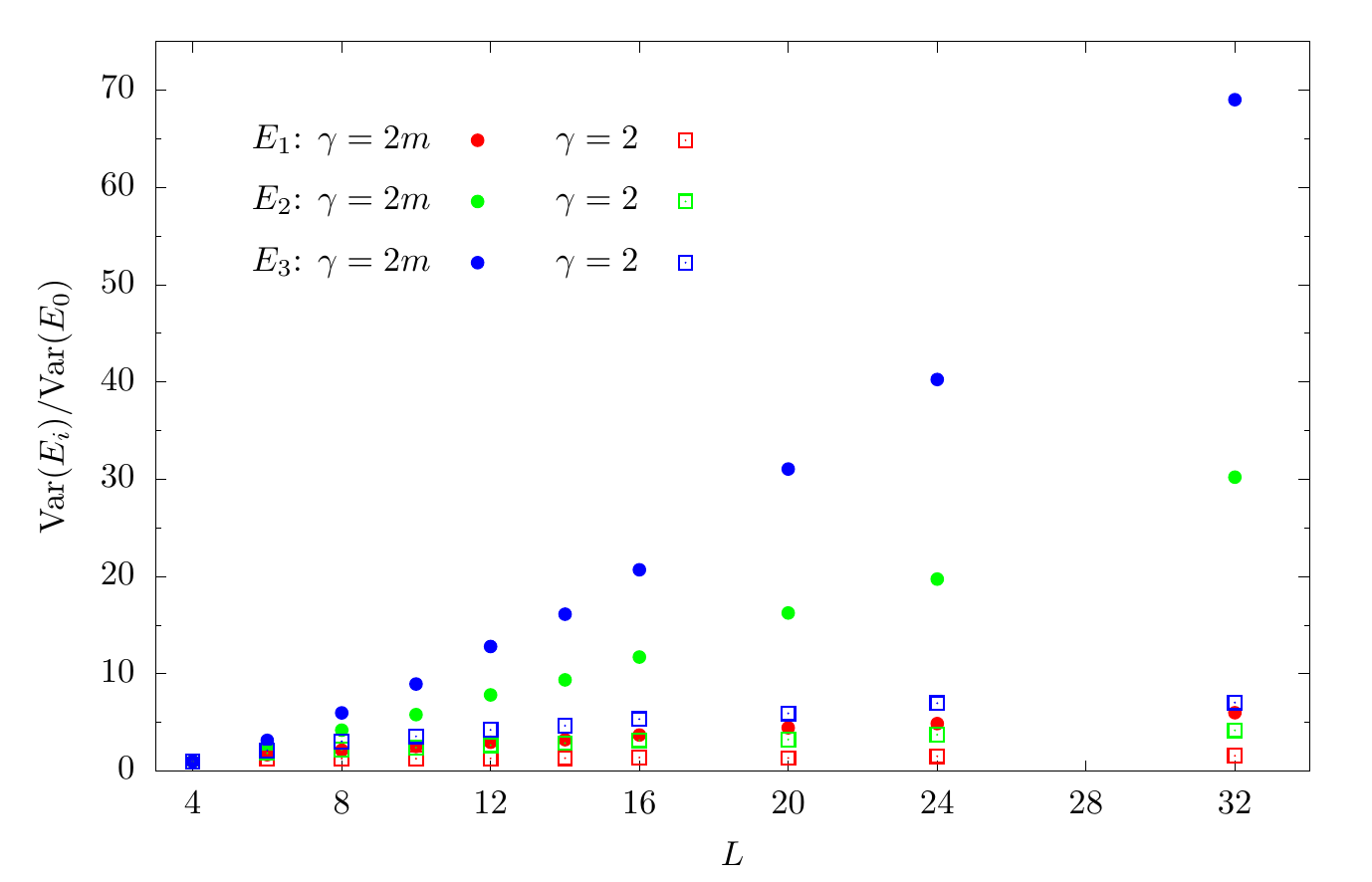}
  \caption{Continuum scaling of the ratios \(\mathop{\rm Var}(E_i)/\mathop{\rm
      Var}(E_0)\) for \(i=1,2,3\) for KSPT, for the cases \(\gamma=2m\) and
    \(\gamma=2\).  The case of \(z=4\) and \(c=0.2\) is shown, and the data are
    normalized at \(L=4\).}
  \label{fig:ScalingVarianceKSPT}
\end{figure}

In Figure~\ref{fig:ScalingVarianceKSPT} we plot the results for the ratios
\(\mathop{\rm Var}(E_i)/\mathop{\rm Var}(E_0)\) as obtained with KSPT.  The two
cases \(\gamma=2m\) and \(\gamma=2\) are shown.  These results are very similar
to those for HSPT: \(\gamma=2m\) leads to larger variances than keeping
\(\gamma=2\) fixed, and these variances grow rapidly with perturbative order as
the continuum limit is approached.

These results show that beyond the lowest perturbative order not only do the
autocorrelations of observables computed using NSPT depend on the parameters of
the algorithms but their variances do too.  This is quite a different situation
to the familiar case of non-perturbative computations.

\subsection{Continuum cost scaling and parameter tuning: the case of KSPT}

\label{subsec:CostScaling}

\subsubsection{Continuum cost scaling} \label{subsubsec:CostScaling}

From the results of the previous subsections it is clear that the most
cost-effective tuning of parameters for a NSPT simulation is not trivial to
determine.  For all cases considered decreasing autocorrelations occurs
concomitantly with increasing variances; the optimal compromise between the two
effects must be found.

\begin{figure}[hptb]
  \centering
  \includegraphics[width=0.70\textwidth]{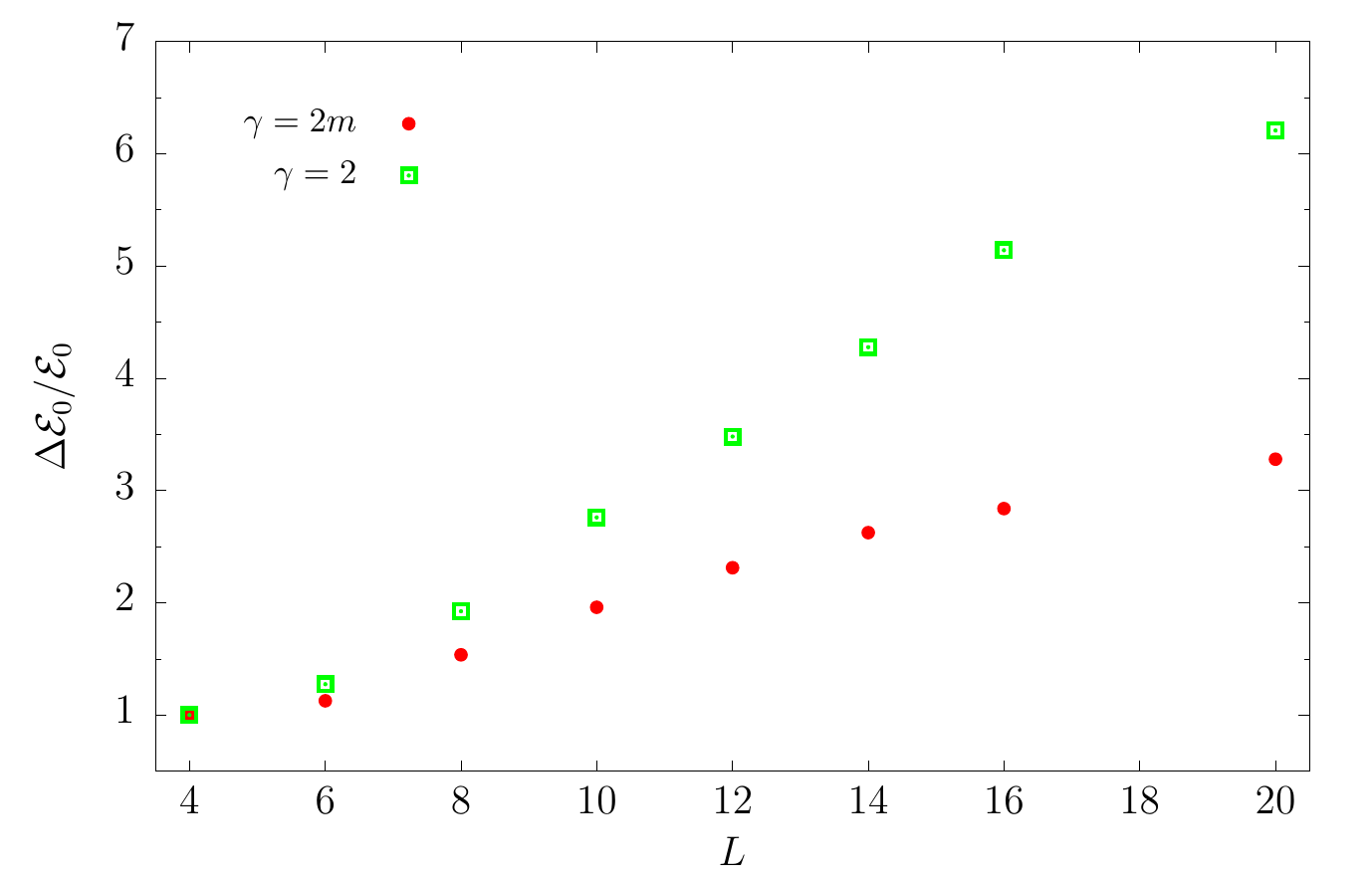}\\
  \includegraphics[width=0.70\textwidth]{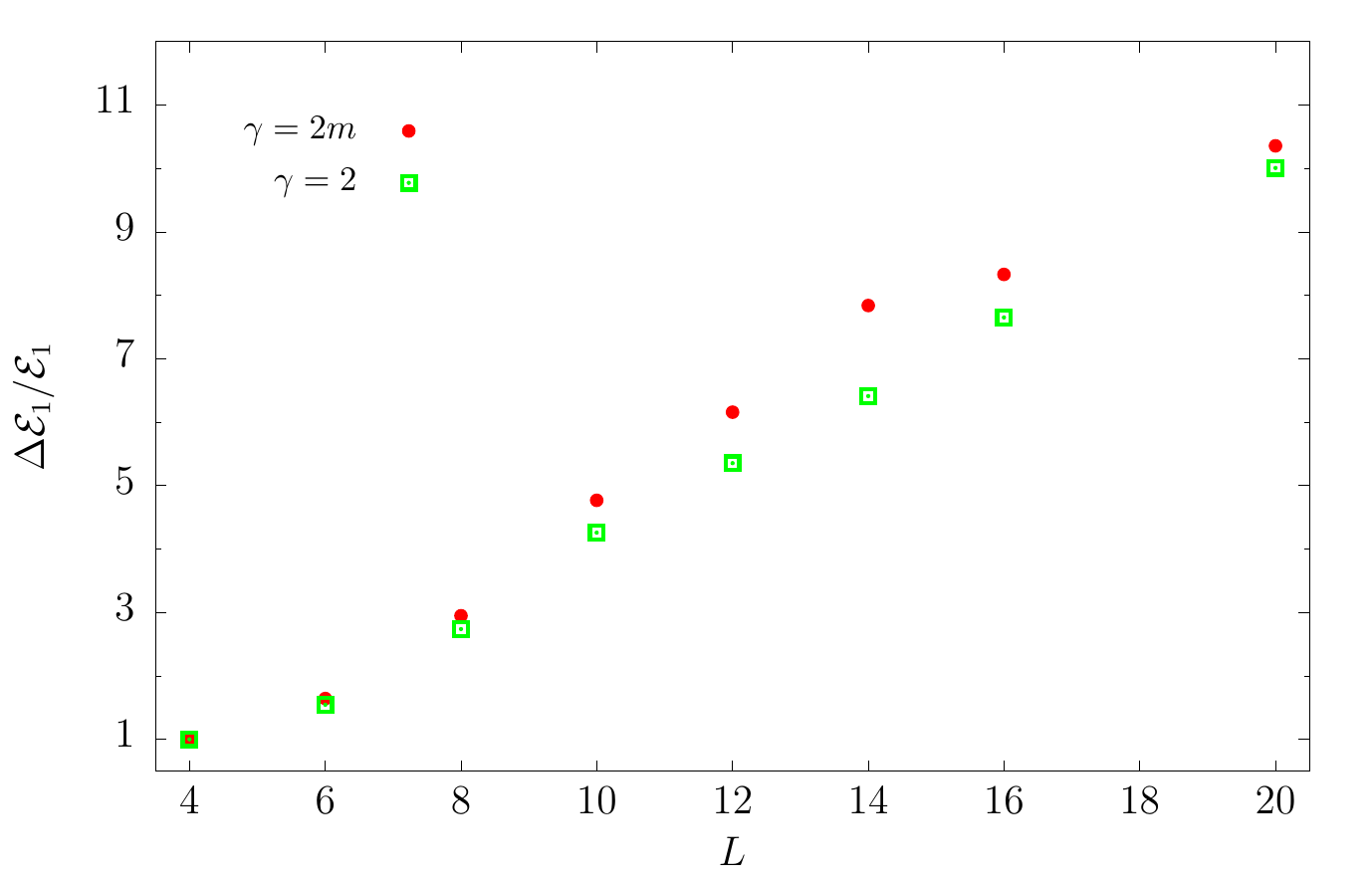}\\
  \includegraphics[width=0.70\textwidth]{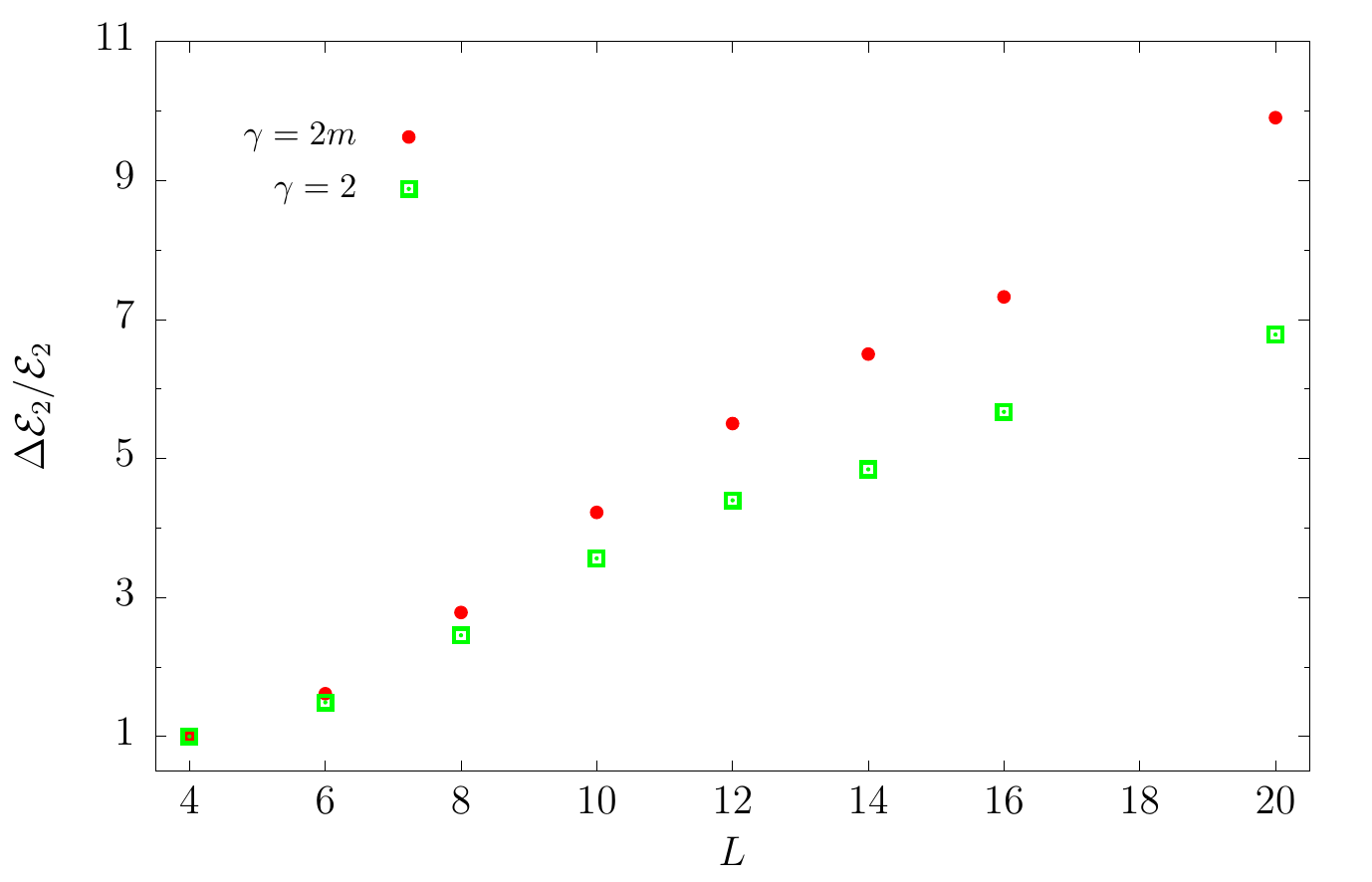}
  \caption{Relative errors \(\Delta \mathcal{E}_i/\mathcal{E}_i\), \(i=0,1,2\),
    as a function of \(L\) for the two cases \(\gamma=2\) and \(\gamma=2m\).
    The data are normalized at \(L=4\).}
  \label{fig:ErrorScalingKSPT}
\end{figure}

The situation is clear if we look directly at the total error
(\ref{eq:ErrorDefinition}) rather than at autocorrelations and variances
separately, and compare the two parameter scalings investigated above.  For
illustration we consider the case of KSPT; HSPT gives very similar results.  In
Figure \ref{fig:ErrorScalingKSPT} we compare the relative error \(\Delta
\mathcal{E}_i/\mathcal{E}_i\) with \(i=0,1,2\) for the cases \(\gamma=2\) and
\(\gamma=2m\).  The number of configurations for the two parameter scalings is
fixed to \(N_{\mbox{\tiny config}}=10^6\) for all the lattice sizes \(4\leq
L\leq20\).  As usual the data is for \(z=4\), \(c=0.2\).  We took \(\delta t =
2/L\) and adjusted the measurement frequency~\(\propto L\).  As expected,
setting \(\gamma=2m\) is beneficial compared to having \(\gamma=2\) at the
lowest perturbative order (top panel of Figure~\ref{fig:ErrorScalingKSPT}).  On
the other hand, when considering higher perturbative orders the case
\(\gamma=2m\) seems to give comparable if not larger errors than fixing
\(\gamma=2\) as~\(L\to\infty\).  Hence, for the range of lattice sizes and
perturbative orders we investigated, the effect of having smaller
autocorrelations for \(\gamma=2m\) appears to be compensated if not overcome by
the corresponding increase of the variances.

\subsubsection{Parameter tuning} \label{subsubsec:ParamTuning}

\begin{figure}[hptb]
  \centering
  \includegraphics[width=0.68\textwidth]{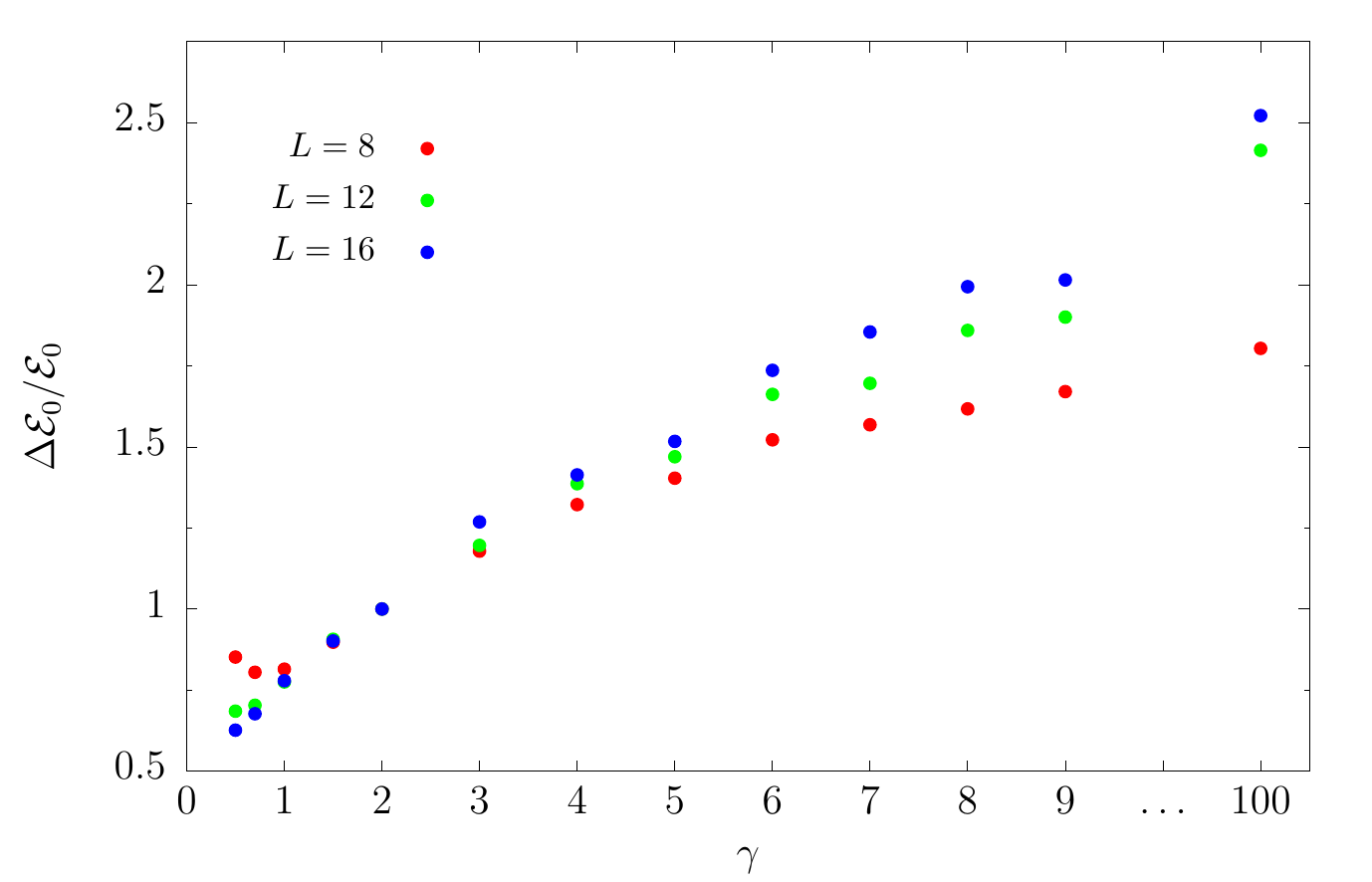}\\
  \includegraphics[width=0.68\textwidth]{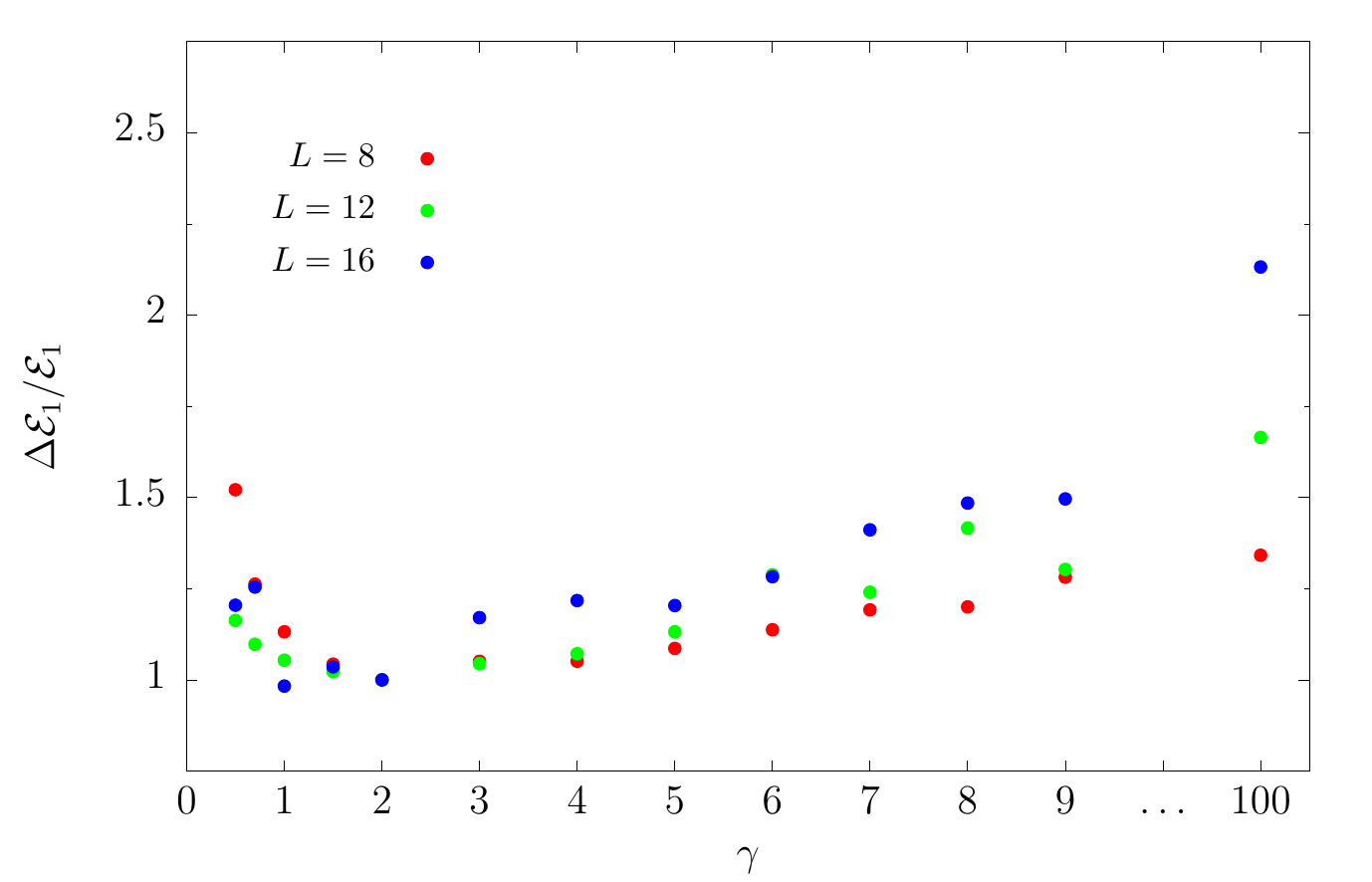}\\
  \includegraphics[width=0.68\textwidth]{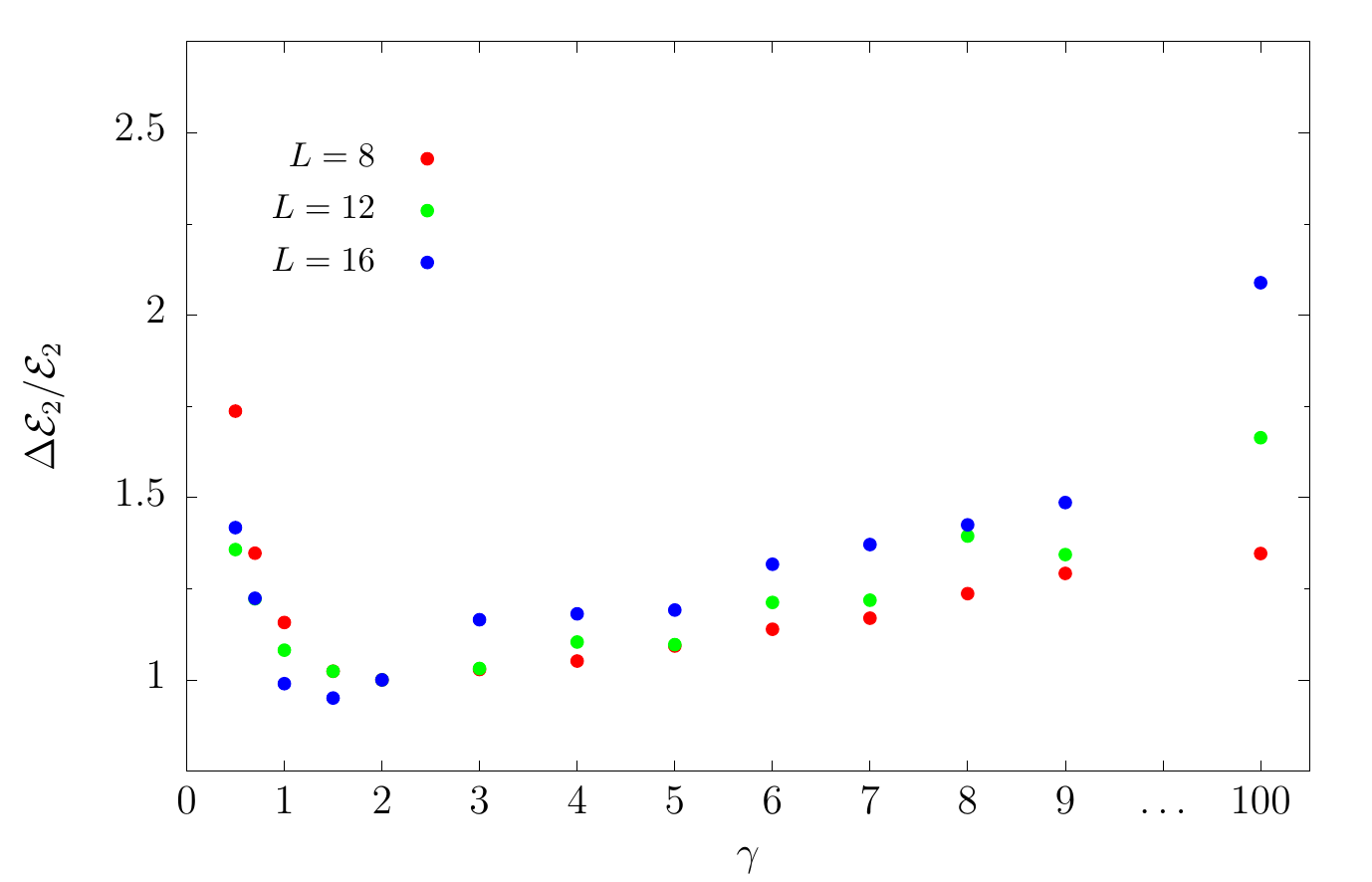}
  \caption{Relative errors \(\Delta \mathcal{E}_i/\mathcal{E}_i\) with
    \(i=0,1,2\) as a function of~\(\gamma\) for \(L=8,12,16\).  The data are
    normalized at~\(\gamma=2\), and the results for \(\gamma=100\) are also
    shown.  At this large value of \(\gamma\) the algorithm is effectively
    integrating the Langevin equation.}
  \label{fig:gammascanKSPT}
\end{figure}

It appears clear that optimizing the performance of the algorithms requires
finding the optimal value of \(\langle\tau\rangle\) or \(\gamma\) for given
lattice parameters, given observables, and the perturbative orders of interest.
Focusing on the case of KSPT again, in Figure \ref{fig:gammascanKSPT} we plot
for example the relative errors \(\Delta \mathcal{E}_i/\mathcal{E}_i\) for
\(i=0,1,2\) as a function of \(\gamma\) for different values of \(L\).  For
each \(L\) and perturbative order, the total number of configurations
\(N_{\mbox{\tiny config}}\) was kept constant as \(\gamma\) was varied, and the
results are normalized by their values at~\(\gamma=2\).  As usual \(z=4\),
\(c=0.2\), and~\(\delta t=2/L\).

At tree-level (top panel) increasing \(\gamma\) leads to an increase of the
relative error except at very small \(\gamma\) values and small lattice size.
This is expected because in this case the variance is independent
of~\(\gamma\), while the autocorrelations increase with \(\gamma\) until they
saturate at some large enough value.  In this regime the algorithm is
effectively integrating the Langevin equation up to step-size errors.  The
situation for the higher perturbative orders \(\mathcal{E}_1\) and
\(\mathcal{E}_2\) is quite different.  For small \(\gamma\) the errors fall
rapidly as \(\gamma\) is increased; as we expect the autocorrelations to be
small in this case we interpret this as a rapid fall of the variances.  For
larger \(\gamma\) values the errors increase only mildly compared to the
situation at tree level.  As at tree-level autocorrelations tend to grow with
\(\gamma\), but this effect is compensated by the variances decreasing as
\(\gamma\) is increased.  In particular, we note that the Langevin limit
\(\gamma\to\infty\) is characterized by having the largest autocorrelations but
the smallest variances.  There is a region of \(\gamma\) values for which the
errors are minimized; in the example considered this does not appear to
strongly depend on either the perturbative order or the lattice size.  This is
comforting as it allows us to tune \(\gamma\) easily and to improve the
efficiency of the algorithm relative to Langevin.

\subsubsection{Cost comparison with LSPT} \label{subsubsec:KSPTvLSPT}

The results of the previous subsection show that a proper tuning of the 
parameter $\gamma$ increases the efficiency of KSPT over its Langevin limit,
$\gamma\to\infty$. In the specific example considered, choosing a value of 
$\gamma\approx2$ appears to be a good compromise for the different perturbative
orders investigated, and it leads to a reduction of the statistical errors at 
fixed cost by a factor $\approx2-2.5$ for $L=16$ as compared to $\gamma\to\infty$;
this corresponds to a factor $\approx4-6$ in the cost at fixed statistical precision.

It is interesting to consider a direct comparison between KSPT and LSPT.
We note that in practice  LSPT differs from KSPT at $\gamma=\infty$ only by 
the different integration scheme used to integrate the Langevin equation. 
Hence, this comparison permits us to quantify the benefits of using efficient
higher-order symplectic integrators in conjunction with a proper tuning of $\gamma$.

To this end, we compared the computational cost for computing the coefficients
${\cal E}_i$, $i=0,\ldots3$, to a specified statistical accuracy using KSPT with
$\gamma=2$ and LSPT. We chose to carry out this comparison with $L=12$, $z=4$ and
$c=0.2$. For $L=12$ the reduction in the cost for a given statistical precision compared
to $\gamma\to\infty$ is a factor $2-3$ for ${\cal E}_k$, $k=1,2,3$, and a 
factor 6 for ${\cal E}_0$ (q.v., Figure~\ref{fig:gammascanKSPT}). 

In view of the results of \S\ref{subsec:tests}, we chose $\delta t=0.5$ for KSPT as 
we expect step-size errors to be very small compared to the precision of this test 
(see below). Similarly, for LSPT we took $\varepsilon=0.01$, which corresponds to the
smallest step-size considered in \S\ref{subsec:tests}. At this value we also expect 
step-size errors to be small, and this is the most expensive of the simulations considered
in the extrapolation $\varepsilon\to0$. At each step we made measurements for both KSPT 
and LSPT, and considered a total of configurations $N_{\rm config}=4\times10^6$ and 
$4\times 10^7$, respectively. The results are collected in Table \ref{tab:KSPTvsLSPT}.

\begin{table}[hbpt]
  \centering
  \begin{tabular}{lllll}
  \toprule
   & ${\cal E}_0\times10^5$ & ${\cal E}_1\times10^8$ & ${\cal E}_2\times10^9$ & 
   ${\cal E}_3\times10^{10}$ \\
  \midrule 
   LSPT & $2.2367(37)$ & $-4.86(13)$ & $2.352(54)$ & $-1.599(40)$ \\
   KSPT & $2.2347(22)$ & $-4.74(12)$ & $2.223(49)$ & $-1.517(35)$ \\
   LPT  & $2.2347$     & $-4.76$     & $2.270$  & $-$ \\
  \bottomrule
  \end{tabular}
 \caption{Results for ${\cal E}_i$, $i=0,\ldots3$ for $L=12$, $z=4$, $c=0.2$
 as obtained using KSPT with $\gamma=2$ and LSPT. We chose $\delta t=0.5$
 for KSPT and $\varepsilon=0.01$ for LSPT, and measured at each step. The total 
 number of configurations generated with the two algorithms is $N_{\rm config}=4\times10^6$
 and $4\times10^7$ for KSPT and LSPT, respectively. The analytic perturbative 
 results (LPT) for ${\cal E}_0$, ${\cal E}_1$, and ${\cal E}_2$ are also given
 for comparison.}
 \label{tab:KSPTvsLSPT}
\end{table}

The KSPT and LSPT results are statistically consistent with each other and with
the closed-form perturbative results (LPT) where these are available. KSPT and
LSPT have approximately the same statistical errors with the numbers of
configurations generated.  The computer time used for updating a \(12^4\) lattice
on a single core of an Intel Xeon E5-2630 Processor (2.4~GHz) is \(0.21s\) for
LSPT and \(0.42s\) for KSPT: this is just the expected ratio of costs between the
RK2 and OMF4 integrators, with the observation that this cost is dominated by
the force computation.%
\footnote{We recall that the RK2 integrator requires three force computations per
	  step whereas the OMF4 integrator requires six (q.v.~\S\ref{subsec:LSPT}
	  and \S\ref{subsec:HSPT}, respectively).}

Thus, after rescaling \(N_{\rm config}\) to have equal statistical
errors, it becomes apparent that KSPT is \(\approx5-7\) times more cost
effective than LSPT is in reaching a given statistical precision on the
higher-order coefficients \({\cal E}_k\), \(k=1,2,3\), and roughly 14 times
more cost effective for \({\cal E}_0\). As KSPT at fixed \(\gamma\) and LSPT
have the same continuum scaling behavior in terms of variances and
autocorrelations, one may expect a similar gain as \(L\to\infty\). Indeed, as
shown in Figure~\ref{fig:gammascanKSPT}, the gain in the statistical errors
appears to become larger for the higher-order fields at larger \(L\); if one
scales the gain in cost accordingly this increases to a factor \(\approx9-10\)
for \({\cal E}_1\) and \({\cal E}_2\). Furthermore, as the continuum limit is
approached, it is advisable to reduce the step-size so as to keep systematic
effects under control, and here again higher-order integrators are more cost
effective.



\section{Conclusions}

NSPT is a powerful technique that permits automation of high order perturbative
computations on the lattice.  As well as providing perturbative lattice
estimates of quantities of interest these methods are interesting for
extracting continuum perturbation theory results in cases where these are
difficult or unfeasible to obtain with continuum perturbative methods.
However, to this end one needs efficient NSPT algorithms in order to be able to
obtain precise results with both systematic and statistical errors under
control.  In particular, such results are desirable for a collection of lattice
resolutions close to the continuum limit so that reliable continuum
extrapolations may be performed.

In this work we investigated some new formulations of NSPT beyond LSPT, with
the goal of finding more cost-effective algorithms.  The first of these
techniques is the recently proposed ISPT~\cite{Luscher:2014mka}.  The first
manifest advantage of this method over standard LSPT is that the results
obtained are exact within statistical errors.  Secondly, the stochastic field
representing the theory to some given order in the couplings is constructed
directly from a set of Gaussian random fields, which are easy to generate.
Despite these attractive features this algorithm has severe limitations beyond
the lowest perturbative orders.  First, similarly to conventional diagrammatic
perturbation theory, the number of diagrams to be computed grows very rapidly
with the perturbative order.  While the cost of evaluating the diagrams is
essentially proportional to the system size, their number increases
exponentially as the perturbative order is increased.  Most importantly, as
shown by the present study, as the continuum limit is approached the
statistical variance of perturbative coefficients computed using ISPT grows
with increasing powers of \(L\) as the perturbative order is increased.
Consequently it appears difficult to extract precise high-order results close
to the continuum limit using this technique.  While the exact details of our
investigation certainly depend on the theory we considered, our conclusions are
not specific to~\(\varphi^4\)-theory.  This has been confirmed by a recent
study in the pure \(\mathop{\rm SU}(3)\) Yang--Mills
theory~\cite{LuscherTalk:2015}, where the nature of the divergences of the
variances was also elucidated.  In summary, the utility of this technique may
be limited to a few low perturbative orders, which can nonetheless be of
interest for some particularly difficult problems.

Although they are not exact the other NSPT algorithms we considered, where the
stochastic fields representing the theory are generated by a Markov chain (or
equivalently a discrete stochastic process), in general have a significantly
better continuum cost scaling than ISPT.  In particular, apart from the
standard LSPT, we considered NSPT based on GHMD algorithms; specifically the
HMD and Kramers algorithms.  With respect to the Langevin implementation, these
allow for a much more accurate discretization of the relevant equations.  This
is so because very efficient high-order symplectic integrators can be employed
for the numerical integration of the MD equations.  With such integrators the
magnitude of the systematic errors is drastically reduced for a given number of
force computations, and in practice one can run these algorithms with a small
enough step-size that step-size extrapolations can be avoided.

As opposed to LSPT, HSPT and KSPT have tunable parameters, the average
trajectory length \(\langle\tau\rangle\) and the amount of partial momentum
refreshment \(\gamma\) respectively, which may be adjusted so as to optimize
their efficiency.  However, beyond the lowest perturbative order finding the
most cost-effective tuning of these parameters is not immediately obvious, in
particular because their optimal continuum scaling is not trivial.  The
situation is complicated by the fact that, unlike the more familiar
non-perturbative simulations, not only do the autocorrelations of the
perturbative coefficients computed in NSPT depend on the parameters of the
chosen algorithm, but so do their variances.  The general trend we observed is
that when an algorithm is tuned to have small autocorrelations, the
corresponding variances tend to increase, and therefore a trade-off between
these two effects must be found.  Moreover, except in the Langevin limit of
these algorithms, analytic understanding of the continuum scaling of both
autocorrelations and variances is missing.

Our analysis indicates that the behavior of the autocorrelations of the high
order fields with respect to the algorithmic parameters is the same as in the
free field case.  The behaviour of the variances is not easily predicted, and
it seems to be different for different perturbative orders.  A consequence of
this is the fact that the optimal parameter scaling suggested by free field
theory is not optimal when higher perturbative orders are considered.  In our
study we did not observe a significant difference in the cost with respect to
the Langevin scaling of the algorithms (\S\ref{subsubsec:CostScaling}).  Finding
the optimal parameter scaling might thus be difficult, as it probably depends
on the details of the calculation considered, i.e., the observables, the
perturbative orders, and range of lattice sizes of interest.

Nonetheless, when investigating the dependence of the errors in KSPT as a
function of~\(\gamma\) (\S\ref{subsubsec:ParamTuning}) we found that for
\(\gamma\approx 2\) the algorithm is significantly better than in its Langevin
limit~\(\gamma\to\infty\), particularly so for large~\(L\).  For example, at
\(L=16\) an improvement by a factor \(\approx4-6\) in the cost of obtaining 
a given statistical precision was observed, depending on the order.  This 
was possible since for the observables studied the optimal value of \(\gamma\)
did not seem to depend much on either \(L\) or the perturbative order. 
When we compared KSPT at $\gamma=2$ with LSPT (\S\ref{subsubsec:KSPTvLSPT}), 
the use of efficient high-order integrators turned out to be beneficial in keeping 
systematic errors under control in a more cost effective way than using 
lower-order Runge-Kutta integrators, keeping this value of $\gamma$  fixed as 
\(L\to\infty\) improves significantly the efficiency of the algorithm over LSPT.
Indeed, although the scaling behaviour of the statistical errors may be the same
one profits from a significantly smaller prefactor, as well as the better scaling
(and prefactor) of the high-order symplectic integrators in controlling step-size errors.

We also observe that HSPT and KSPT have similar performance: for \(\langle
\tau\rangle = C/\gamma\) with \(C=O(1)\) the two algorithms have comparable
autocorrelations in molecular dynamics units, and comparable variances.

In conclusion, the novel NSPT methods presented here offer a simple and natural
development from the standard Langevin-based algorithms.  In particular, we
have provided evidence that they can significantly improve on previous methods
hence allowing more precise results.  Of course, a natural follow-up of our
study is to consider the application of these techniques to a realistic problem
in order to determine whether the improvement provided by HSPT or KSPT is
significant in practice.  These methods have been used and are under further
development for the more interesting case of gauge
theories~\cite{DallaBrida:2016dai,DallaBrida:2017tru}.



\section*{Acknowledgments}

M.D.B. especially thanks Martin L\"uscher for the pleasant and fruitful
collaboration in further understanding and developing NSPT.  He also thanks
Chris Monahan and Ulli Wolff for interesting discussions, and he is
grateful to CERN for hospitality and support.  A.D.K. and M.G. are funded by STFC
Consolidated Grant No. ST/J000329/1.

We thank the computer center at DESY--Zeuthen for computer resources and
support, and the University of Edinburgh for use of the ECDF cluster~(eddie).


%
\cleardoublepage
\appendix

\section{Renormalization procedure}
\label{sec:appA}

\subsection{Coupling renormalization}
\label{subsec:CouplingRenormalization}

In regularized \(\phi^4\) theory we may compute an observable \(\obs\) as a
perturbative expansion in the bare coupling~\(g_0\).  However, in order to take
the continuum limit of its expectation value, it is first of all necessary to
express this perturbative series in terms of a renormalized coupling~\(g\).  Of
course, at finite lattice cutoff, the two are entirely equivalent as formal
expansions and may readily be transformed into each other.

Suppose we have computed the perturbative expansion of the renormalized
coupling \(g\) as a power series in the bare coupling~\(g_0\),
\begin{equation}
  g = g_0 + \sum_{k\geq2} c_k g_0^k.
  \label{eq:g-expansion}
\end{equation}
We may then \emph{revert\/} the expansion of \(g\) in terms of~\(g_0\) by
writing (\ref{eq:g-expansion})~as
\begin{equation}
  g_0 = g - \sum_{k\geq2} c_k g_0^k,
  \label{eq:start}
\end{equation}
and then recursively substituting (\ref{eq:start}) into itself to obtain
\begin{align}
  g_0 & = g - \sum_{k\geq2} c_k \Bigl(g - \sum_{\ell\geq2} c_\ell g_0^\ell\Bigr)^k
    \nonumber \\
  & = g - c_2 g^2 + (2c_2^2 - c_3) g^3 + (-5c_2^3 + 5c_2 c_3 - c_4) g^4
    \nonumber \\
  & \qquad + (14 c_2^4 - 21c_2^2 c_3 + 6c_2 c_4 + 3c_3^2 - c_5) g^5 \nonumber \\
  & \qquad + (-42 c_2^5 + 84 c_2^3 c_3 - 28 c_2^2 c_4 - 28 c_2 c_3^2
    + 7 c_2 c_5 + 7 c_3 c_4 - c_6) g^6 + \cdots,
  \label{eq:reverted_series}
\end{align}
noting that \(\order(g_0^N) = \order(g^N)\).

Suppose that we have also computed the expansion of some operator of interest
\(\obs\) in powers of~\(g_0\)
\begin{equation}
  \obs = \sum_{k\geq0} \obs_k g_0^k,
  \label{eq:op-g-expansion}
\end{equation}
Then by substituting (\ref{eq:reverted_series}) into (\ref{eq:op-g-expansion})
we obtain an expression for the expansion of \(\obs\) in powers of~\(g\):
\begin{align*}
  \obs & = \obs_0 + \obs_1 g + (-c_2 \obs_1 + \obs_2) g^2
    + \bigl((2 c_2^2 - c_3) \obs_1 - 2 c_2 \obs_2 + \obs_3\bigr) g^3 \\
  & \qquad + \bigl((-5 c_2^3 + 5 c_2 c_3 - c_4) \obs_1
    + (5 c_2^2 - 2 c_3) \obs_2 - 3 c_2 \obs_3 + \obs_4\bigr) g^4 \\
  & \qquad + \bigl((14 c_2^4 - 21 c_2^2 c_3 + 6 c_2 c_4 + 3 c_3^2 - c_5) \obs_1
    + (-14 c_2^3 + 12 c_2 c_3 - 2 c_4) \obs_2 \\
  & \qquad\qquad + (9 c_2^2 - 3 c_3) \obs_3
    - 4 c_2 \obs_4 + \obs_5\bigr) g^5 \\
  & \qquad + \bigl((-42 c_2^5 + 84 c_2^3 c_3 - 28 c_2^2 c_4 - 28 c_2 c_3^2
    + 7 c_2 c_5 + 7 c_3 c_4 - c_6) \obs_1 \\
  & \qquad\qquad + (42 c_2^4 - 56 c_2^2 c_3 + 14 c_2 c_4 + 7 c_3^2 - 2 c_5)
    \obs_2 \\
  & \qquad\qquad + (-28 c_2^3 + 21 c_2 c_3 - 3 c_4) \obs_3
    + (14 c_2^2 - 4 c_3) \obs_4 - 5 c_2 \obs_5 + \obs_6\bigr) g^6 + \cdots
\end{align*}
For the numerical computation of the perturbative expansion of \(\obs\) we are
therefore free to consider an expansion in powers of \(g_0\) as this is
entirely equivalent --- as formal power series --- to expansion in powers
of~\(g\).

\subsection{Mass renormalization}
\label{subsec:MassRenormalization}

The stochastic field \(\phi\) is considered to be of the form
\begin{equation}
  \phi(x)=\sum_{k,l\geq0} \phi_{k,\ell}(x) g_0^k (\delta m^2)^\ell
\end{equation}
where \(g_0\) is the bare coupling and \(\delta m^2\) is the mass
counterterm.\footnote{Remember that \(\delta m^2\) has contributions of order
  \(g_0^n\) for \(n\geq1\) when it has been determined from the renormalization
  conditions (see the following discussion).}  Once the table of numbers
\(\phi_{k,\ell}\) has been computed, the expectation value \(\langle\cdots
\rangle_\eta\) of functions of these quantities may be estimated, but they must
be fitted to the renormalization conditions in order to compute physical
quantities.  Here we shall present algebraic expressions for the formal power
series manipulation in order to explain the renormalization procedure; in
actual computations we automated these formal manipulations using the numerical
values of the coefficients.

The renormalization condition (\ref{eq:MassDefinition}) that defines~\(m^2\)
can be rewritten~as
\[
  m^2 = \hat p_*^2 \frac{\chi_2^*}{\chi_2 - \chi_2^*}.
\]
Therefore, since we can calculate \(\chi_2\) and \(\chi_2^*\) as power series
in both \(g_0\) and~\(\delta m^2\) 
\begin{align*}
  \chi_2 &= \Bigl\langle\tilde\phi_{0,0}(0)^2\Bigr\rangle
      + 2\Bigl\langle\tilde\phi_{0,0}(0) \tilde\phi_{0,1}(0)\Bigr\rangle\delta m^2
    + \Bigl\langle2\tilde\phi_{0,0}(0) \tilde\phi_{0,2}(0)
      + \tilde\phi_{0,1}(0)^2\Bigr\rangle \delta m^4 \\
  &\qquad + 2\Bigl\langle\tilde\phi_{0,0}(0) \tilde\phi_{0,3}(0)
      + \tilde\phi_{0,1}(0)\tilde\phi_{0,2}(0)\Bigr\rangle \delta m^6 \\
  &\quad + 2\biggl(\Bigl\langle\tilde\phi_{0,0}(0) \tilde\phi_{1,0}(0)\Bigr\rangle
    + \Bigl\langle\tilde\phi_{0,0}(0) \tilde\phi_{1,1}(0)
      + \tilde\phi_{1,0}(0) \tilde\phi_{0,1}(0)\Bigr\rangle\delta m^2 \\
  &\qquad + \Bigl\langle\tilde\phi_{0,0}(0) \tilde\phi_{1,2}(0)
      + \tilde\phi_{1,0}(0) \tilde\phi_{0,2}(0)
      + \tilde\phi_{0,1}(0) \tilde\phi_{1,1}(0)\Bigr\rangle\delta m^4\biggr) g_0 \\
  &\quad + \biggl(\Bigl\langle2\tilde\phi_{0,0}(0) \tilde\phi_{2,0}(0)
      + \tilde\phi_{1,0}(0)^2\Bigr\rangle \\
  &\qquad + 2\Bigl\langle\tilde\phi_{0,0}(0) \tilde\phi_{2,1}(0)
      + \tilde\phi_{1,0}(0) \tilde\phi_{1,1}(0)
      + \tilde\phi_{0,1}(0) \tilde\phi_{2,0}(0)\Bigr\rangle\delta m^2\biggr) g_0^2 \\
  &\quad + 2\Bigl\langle\tilde\phi_{0,0}(0) \tilde\phi_{3,0}(0)
      + \tilde\phi_{1,0}(0) \tilde\phi_{2,0}(0)\Bigr\rangle g_0^3 + \order(g_0^4)
\end{align*}

\begin{align*}
  \chi_2^* & = \Bigl\langle\tilde\phi_{0,0}(p_*) \tilde\phi_{0,0}(-p_*)\Bigr\rangle
  + \Bigl\langle\tilde\phi_{0,0}(p_*) \tilde\phi_{0,1}(-p_*)
    + \tilde\phi_{0,0}(-p_*) \tilde\phi_{0,1}(p_*)\Bigr\rangle \delta m^2 \\
  &\qquad + \Bigl\langle\tilde\phi_{0,0}(p_*) \tilde\phi_{0,2}(-p_*)
    + \tilde\phi_{0,1}(p_*) \tilde\phi_{0,1}(-p_*)
    + \tilde\phi_{0,0}(-p_*) \tilde\phi_{0,2}(p_*)\Bigr\rangle \delta m^4 \\
  &\qquad + \Bigl\langle\tilde\phi_{0,0}(p_*) \tilde\phi_{0,3}(-p_*)
    + \tilde\phi_{0,1}(p_*) \tilde\phi_{0,2}(-p_*) \\
  &\qquad\quad + \tilde\phi_{0,2}(p_*) \tilde\phi_{0,1}(-p_*)
    + \tilde\phi_{0,0}(-p_*) \tilde\phi_{0,3}(p_*)\Bigr\rangle \delta m^6 \\
  &\quad + \biggl(\Bigl\langle\tilde\phi_{0,0}(p_*) \tilde\phi_{1,0}(-p_*)
    + \tilde\phi_{0,0}(-p_*) \tilde\phi_{1,0}(p_*)\Bigr\rangle \\
  &\qquad + \Bigl\langle\tilde\phi_{0,0}(p_*) \tilde\phi_{1,1}(-p_*)
    + \tilde\phi_{1,0}(p_*) \tilde\phi_{0,1}(-p_*) \\
  &\qquad\quad + \tilde\phi_{0,1}(p_*) \tilde\phi_{1,0}(-p_*)
    + \tilde\phi_{0,0}(-p_*) \tilde\phi_{1,1}(p_*)\Bigr\rangle \delta m^2 \\
  &\qquad + \Bigl\langle\tilde\phi_{0,0}(p_*) \tilde\phi_{1,2}(-p_*)
    + \tilde\phi_{1,0}(p_*) \tilde\phi_{0,2}(-p_*) \\
  &\qquad\quad + \tilde\phi_{0,1}(p_*) \tilde\phi_{1,1}(-p_*)
    + \tilde\phi_{1,1}(p_*) \tilde\phi_{0,1}(-p_*) \\
  &\qquad\quad + \tilde\phi_{0,2}(p_*) \tilde\phi_{1,0}(-p_*)
    + \tilde\phi_{0,0}(-p_*) \tilde\phi_{1,2}(p_*)\Bigr\rangle \delta m^4\biggr) g_0 \\
  &\quad + \biggl(\Bigl\langle\tilde\phi_{0,0}(p_*) \tilde\phi_{2,0}(-p_*)
    + \tilde\phi_{1,0}(p_*) \tilde\phi_{1,0}(-p_*)
    + \tilde\phi_{0,0}(-p_*) \tilde\phi_{2,0}(p_*)\Bigr\rangle \\
  &\qquad + \Bigl\langle\tilde\phi_{0,0}(p_*) \tilde\phi_{2,1}(-p_*)
    + \tilde\phi_{1,0}(p_*) \tilde\phi_{1,1}(-p_*) \\
  &\qquad\quad + \tilde\phi_{0,1}(p_*) \tilde\phi_{2,0}(-p_*)
    + \tilde\phi_{2,0}(p_*) \tilde\phi_{0,1}(-p_*) \\
  &\qquad\quad + \tilde\phi_{1,1}(p_*) \tilde\phi_{1,0}(-p_*)
    + \tilde\phi_{0,0}(-p_*) \tilde\phi_{2,1}(p_*)\Bigr\rangle \delta m^2\biggr) g_0^2 \\
  &\quad + \biggl(\Bigl\langle\tilde\phi_{0,0}(p_*) \tilde\phi_{3,0}(-p_*)
    + \tilde\phi_{1,0}(p_*) \tilde\phi_{2,0}(-p_*) \\
  &\qquad + \tilde\phi_{2,0}(p_*) \tilde\phi_{1,0}(-p_*)
    + \tilde\phi_{0,0}(-p_*) \tilde\phi_{3,0}(p_*)\Bigr\rangle\biggr) g_0^3 + O(g_0^4)
\end{align*}
where we defined the Fourier transform of the coefficient fields as,
\begin{equation}
 \tilde\phi_{k,l}(p)=\frac{1}{L^2}\sum_{x\in \Omega }e^{-ipx} \phi_{k,l}(x),\quad
 p\in \tilde \Omega,
 \nonumber
\end{equation}
we can multiply and invert\footnote{The \emph{inverse\/} of a power series
  \(S(g_0,\delta m)\) is the power series for \(1/S(g_0,\delta m)\).}  \(\chi\)
and \(\chi^*\) to compute \(m^2\) as power series in \(g_0\) and~\(\delta m^2\)
\begin{equation}
  m^2 = \sum_{k,\ell\geq0} a_{k,\ell}\,g_0^k(\delta m^2)^\ell
  \label{eq:MassRenormalization2}
\end{equation}
where the coefficients \(a_{k,\ell}\)~are
\begin{align*}
  a_{0,0}
    & = \frac{\Bigl\langle\tilde\phi_{0,0}(p_*)\tilde\phi_{0,0}(-p_*)\Bigr\rangle}
      {\Bigl\langle\tilde\phi_{0,0}(0)^2 - \tilde\phi_{0,0}(p_*)\tilde\phi_{0,0}(-p_*)\Bigr\rangle}
      \,\hat p_*^2, \\[1ex]
  a_{1,0} & = -\frac{\begin{array}{l}
    2\Bigl\langle\tilde\phi_{0,0}(0)\tilde\phi_{1,0}(0)\Bigr\rangle
      \Bigl\langle\tilde\phi_{0,0}(p_*)\tilde\phi_{0,0}(-p_*)\Bigr\rangle \\
    \qquad\qquad - \Bigl\langle\tilde\phi_{0,0}(0)^2\Bigr\rangle
      \Bigl\langle\tilde\phi_{0,0}(p_*)\tilde\phi_{1,0}(-p_*)
        + \tilde\phi_{0,0}(-p_*)\tilde\phi_{1,0}(p_*)\Bigr\rangle
    \end{array}}
    {\Bigl\langle\tilde\phi_{0,0}(0)^2
      - \tilde\phi_{0,0}(p_*)\tilde\phi_{0,0}(-p_*)\Bigr\rangle^2} \,\hat p_*^2, \\[1ex]
  a_{0,1} & = -\frac{\begin{array}{l}
    2\Bigl\langle\tilde\phi_{0,0}(0)\tilde\phi_{0,1}(0)\Bigr\rangle
      \Bigl\langle\tilde\phi_{0,0}(p_*)\tilde\phi_{0,0}(-p_*)\Bigr\rangle \\
    \qquad\qquad - \Bigl\langle\tilde\phi_{0,0}(0)^2\Bigr\rangle
      \Bigl\langle\tilde\phi_{0,0}(p_*)\tilde\phi_{0,1}(-p_*)
        + \tilde\phi_{0,0}(-p_*)\tilde\phi_{0,1}(p_*)\Bigr\rangle
    \end{array}}
    {\Bigl\langle\tilde\phi_{0,0}(0)^2
      - \tilde\phi_{0,0}(p_*)\tilde\phi_{0,0}(-p_*)\Bigr\rangle^2} \,\hat p_*^2,
\end{align*}
and so forth.  By construction \(a_{0,0} = m^2\), so at lowest order in \(g_0\)
the mass \(m\) is the mass that enters the scalar
propagator~(\ref{eq:ScalarPropagator}).  Having determined the coefficient
\(a_{k,\ell}\) in (\ref{eq:MassRenormalization2}) we can now determine the
coefficients \(m_k^2\) of the expansion
\[
  \delta m^2 = \sum_{k\geq1} m^2_k\,g_0^k
\]
by imposing the relation (\ref{eq:MassRenormalization2}) order by order in
\(g_0\), thus obtaining
\begin{align*}
  m_1^2 &= -\frac{a_{1,0}}{a_{0,1}}, \\[1ex]
  m_2^2 &= -\frac{a_{2,0} + a_{1,1} m_1^2 + a_{0,2} m_1^4}{a_{0,1}}, \\[1ex]
  m_3^2 &= -\frac{a_{3,0} + a_{2,1} m_1^2 + 2 a_{0,2} m_1^2 m_2^2 + a_{1,2} m_1^4
    + a_{1,1} m_2^2 + a_{0,3} m_1^6}{a_{0,1}}, \\[1ex]
  m_4^2 &= -\frac{\begin{array}{l}
    a_{4,0} + a_{3,1} m_1^2 + a_{2,1} m_2^2 + a_{1,1} m_3^2
      + a_{2,2} m_1^4 + 2 a_{1,2} m_1^2 m_2^2 \\
    \qquad\qquad  + a_{0,2} m_2^4 + 2 a_{0,2} m_1^2 m_3^2
      + a_{1,3} m_1^6 + 3 a_{0,3} m_1^4 m_2^2 + a_{0,4} m_1^8 
    \end{array}}{a_{0,1}}, \\
  &\vdots
\end{align*}
Once \(\delta m^2\) is determined, the field \(\phi\) and any other observable
previously computed as a series in \(\delta m^2\) and \(g_0\) can be reduced to
a series in \(g_0\) alone.

\subsection{Wavefunction renormalization}

The renormalization of a generic correlation function by the wavefunction
renormalization, or any multiplicative renormalization factor, does not present
any additional difficulty.  We may compute \(Z\) as a power series in \(g_0\)
and \(\delta m^2\) from the renormalization condition~(\ref{eq:Z_def}),
\[
  Z = m^2\chi = \sum_{k,\ell\geq0} Z_{k,\ell}\,g_0^k(\delta m^2)^\ell,
\]
where
\begin{align*}
  Z_{0,0} &= a_{0,0}\Bigl\langle\tilde\phi_{0,0}(0)^2\Bigr\rangle \\[2ex]
  Z_{1,0} &= a_{1,0}\Bigl\langle\tilde\phi_{0,0}(0)^2\Bigr\rangle
    + 2a_{0,0}\Bigl\langle\tilde\phi_{1,0}(0)\tilde\phi_{0,0}(0)\Bigr\rangle \\[2ex]
  Z_{0,1} &= a_{0,1}\Bigl\langle\tilde\phi_{0,0}(0)^2\Bigr\rangle
  + 2a_{0,0}\Bigl\langle\tilde\phi_{0,0}(0)\tilde\phi_{0,1}(0)\Bigr\rangle \\[2ex]
  Z_{2,0} &= 2a_{0,0} \Bigl\langle\tilde\phi_{0,0}(0) \tilde\phi_{2,0}(0)\Bigr\rangle
    + a_{0,0} \Bigl\langle\tilde\phi_{1,0}(0)^2)\Bigr\rangle \\
  &\qquad + 2a_{1,0} \Bigl\langle\tilde\phi_{0,0}(0) \tilde\phi_{1,0}(0)\Bigr\rangle
    + a_{2,0} \Bigl\langle\tilde\phi_{0,0}(0)^2)\Bigr\rangle \\[2ex]
  Z_{1,1} &= 2a_{0,0} \Bigl\langle\tilde\phi_{0,0}(0) \tilde\phi_{1,1}(0)\Bigr\rangle
    + 2a_{0,0} \Bigl\langle\tilde\phi_{1,0}(0) \tilde\phi_{0,1}(0)\Bigr\rangle \\
  &\qquad + 2a_{0,1} \Bigl\langle\tilde\phi_{0,0}(0) \tilde\phi_{1,0}(0)\Bigr\rangle
    + 2a_{1,0} \Bigl\langle\tilde\phi_{0,0}(0) \tilde\phi_{0,1}(0)\Bigr\rangle
    + a_{1,1} \Bigl\langle\tilde\phi_{0,0}(0)^2\Bigr\rangle \\[2ex]
  Z_{0,2} &= 2a_{0,0} \Bigl\langle\tilde\phi_{0,0}(0) \tilde\phi_{0,2}(0)\Bigr\rangle
    + a_{0,0} \Bigl\langle\tilde\phi_{0,1}(0)^2\Bigr\rangle \\
  &\qquad + 2a_{0,1} \Bigl\langle\tilde\phi_{0,0}(0) \tilde\phi_{0,1}(0)\Bigr\rangle
    + a_{0,2} \Bigl\langle\tilde\phi_{0,0}(0)^2\Bigr\rangle,
\end{align*}
and so forth.  We can now compute a renormalized correlation function as a
power series in \(g_0\) and the renormalized mass \(m\) as
\begin{align*}
  & Z^{n/2} \bigl\langle\phi(x_1)\cdots\phi(x_n)\bigr\rangle
  = \bigg(\sum_{k,\ell\geq0} Z_{k,\ell}\,g_0^k(\delta m^2)^\ell\bigg)^{n/2} \\
  & \qquad \times
    \left\langle\sum_{k_1,\ell_1\geq0}\phi_{k_1,\ell_1}(x_1)g_0^{k_1}
      (\delta m^2)^{\ell_1} \:\cdots
    \sum_{k_n,\ell_n\geq0}\phi_{k_n,\ell_n}(x_n)g_0^{k_n}
      (\delta m^2)^{\ell_n}\right\rangle \\
  &= \bigg(\sum_{k,l\geq0} Z_{k,\ell}\,g_0^k
    \Big(\sum_{j\geq1} m_j^2\,g_0^j\Big)^\ell\bigg)^{n/2} \\
  &\qquad \times \left\langle\sum_{k_1,\ell_1\geq0}\phi_{k_1,\ell_1}(x_1)g_0^{k_1}
    \Big(\sum_{j_1\geq1} m_{j_1}^2\,g_0^{j_1}\Big)^{\ell_1} \cdots
    \sum_{k_n,\ell_n\geq0}\phi_{k_n,\ell_n}(x_n)g_0^{k_n}\Big(\sum_{j_n\geq1}
    m_{j_n}^2\,g_0^{j_n}\Big)^{\ell_n}\right\rangle.
\end{align*}
This expansion in the bare coupling \(g_0\) can be replaced by one in the
renormalized coupling \(g\) as explained in
\S\ref{subsec:CouplingRenormalization}, and the correlation function is then
properly renormalized.


%

\cleardoublepage

\bibliographystyle{JHEP}
\bibliography{biblio}

\end{document}